\documentclass[twocolumn,twocolappendix]{aastex631}

\usepackage{graphicx}
\usepackage{subfigure}
\usepackage{verbatim}
\usepackage{makecell}
\usepackage{multirow}

\usepackage[acronym,nohypertypes={acronym},nomain]{glossaries}

\newacronym{BNS}{BNS}{Binary Neutron Star}
\newacronym{Mtov}{M$_{\text{TOV}}$}{Tolman–Oppenheimer–Volkoff limit}
\newacronym{LIGO}{LIGO}{Laser Interferometer Gravitational-wave Observatory}
\newacronym{SNR}{SNR}{Signal-to-Noise ratio}
\newacronym{GRB}{GRB}{Gamma-Ray Burst}
\newacronym{GW}{GW}{Gravitational Wave}
\newacronym{VLA}{VLA}{Very Large Array}
\newacronym{omega}{$\Omega$}{sky localization uncertainty at 90\% confidence}
\newacronym{z}{z}{redshift}
\newacronym{EoS}{EoS}{Equation of State}
\newacronym{TOV}{TOV}{Tolman-Oppenheimer-Volkoff}
\newacronym{LVK}{LVK}{LIGO-Virgo-KAGRA}
\newacronym{ET}{ET}{Einstein Telescope}
\newacronym{CE}{CE}{Cosmic Explorer}
\newacronym{ngVLA}{ngVLA}{next generation Very Large Array}

\newacronym{4020ET}{CE40-CE20-ET}{CE40-CE20-ET}
\newacronym{40ETA}{CE40-ET-A}{CE40-ET-A}
\newacronym{40ET}{CE40-ET}{CE40-ET}
\newacronym{4020}{CE40-CE20}{CE40-CE20}
\newacronym{4020A}{CE40-CE20-A}{CE40-CE20-A}
\newacronym{HLA}{H-L-A}{Hanford-Livingston-India}
\newacronym{MPSAC}{MPSAC}{Mathematical and Physical Sciences Advisory Committee}
\newacronym{O5}{O5}{the fifth observing run of Advanced LIGO, Virgo, and KAGRA}
\begin{document}

\title{Probing Binary Neutron Star Merger Ejecta and Remnants with Gravitational Wave and Radio Observations}

\correspondingauthor{Kara Merfeld}
\email{kmerfel1@jh.edu}

\author[0000-0003-1773-5372]{Kara Merfeld}
\affiliation{William H. Miller III Department of Physics and Astronomy, Johns Hopkins University, Baltimore, MD 21218, USA}

\author[0000-0001-8104-3536]{Alessandra Corsi}
\affiliation{William H. Miller III Department of Physics and Astronomy, Johns Hopkins University, Baltimore, MD 21218, USA}

\begin{abstract}
We present a study aimed at quantifying the detectability of radio counterparts of binary neutron star (BNS) mergers with total masses $\lesssim 3$\,M$_{\odot}$, which may form neutron star (NS) remnants. We focus on mergers localized by gravitational-wave (GW) observations to sky areas $\lesssim 10$\,deg$^2$, a precision that greatly facilitates optical counterpart identification and enables radio discovery even without detections at other wavelengths. Widely separated GW detectors are essential for building samples of well-localized BNS mergers accessible to US-based radio telescopes, with minimum yearly detection rates (assuming the smallest values of the BNS local merger rate)  ranging from a few with current GW detectors to hundreds with next-generation GW instruments. Current GW networks limit well-localized detections to $z\lesssim 0.2$, while next-generation GW detectors extend the reach to $z\lesssim 0.8$, encompassing the median redshift of short gamma-ray bursts (GRBs). With next-generation radio arrays operating at a several tens of GHz and providing an order of magnitude improvement in sensitivity compared to the most sensitive ones available today, short GRB-like jet afterglows can be detected for a large fraction of the considered BNS mergers. At lower radio frequencies, detections with current radio interferometric arrays are feasible, though subject to synchrotron self-absorption effects. The enhanced sensitivity and survey speed of future radio interferometers operating at a few GHz, combined with the higher detection rate of well-localized BNSs enabled by next-generation GW observatories, are key to probing disk-wind and dynamical ejecta afterglows, as well as remnant diversity. 
\end{abstract}


\section{Introduction} \label{sec:intro}
GW170817 was the first \gls{BNS} merger to be discovered in gravitational waves (GWs) by the LIGO and Virgo detectors \citep{Abbott_2017_GW170817}. As of today, GW170817 remains the only \gls{BNS} merger directly associated with a gamma-ray burst (GRB) and with a counterpart detected in practically all bands of the electromagnetic (EM)  spectrum \citep{Smartt_2017,Abbott_2017_MM,Hallinan_2017,Resmi_2018,Troja_2017,Andreoni_2017,Balasubramanian_2021,Balasubramanian_2022,Troja_2020,Alexander_2017,Valenti_2017,Soares_Santos_2017,Villar_2017,Nynka_2018,Margutti_2018,Troja_2019,Lamb_2019,Hajela_2019,Evans_2017,Dobie_2018,Alexander_2018,Makhathini_2021,Mooley_2018}. While a lot was learned from the case of GW170817, the multi-messenger population properties of BNSs remain largely unexplored.

Probing a large population of BNSs via GW and EM observations in general, and via GW plus radio observations in particular, is key to answering several questions left open by GW170817 \citep[e.g., ][and references therein]{Corsi2024a,Corsi2024b}, including one that is key to the study presented here, namely, what is the nature of the merger remnant. Indeed, 
the fate of the remnant of a \gls{BNS} merger is highly dependent on the properties of the BNS system, and particularly of its total mass given a certain \gls{EoS} of nuclear matter. A stable neutron star (NS) can form if the mass of the merger remnant does not exceed the \gls{TOV} limit for the maximum stable mass of non-rotating NSs \citep[e.g.,][]{PhysRev.55.374, PhysRev.55.364,Fan_2024,Margalit_2017,das2023comparativestudymaximummass,Han_2023}. A remnant with mass exceeding the \gls{TOV} limit (hereafter, $M_{\rm TOV}$) could still form a short-lived, hypermassive NS supported by differential rotation, or a longer-lived supramassive NS supported by uniform rotation \citep{Sarin_2021,Baumgarte_2000}. 

While NSs are likely to be a rare outcome of BNS mergers, next generation GW detectors, such as Cosmic Explorer \citep[CE;][]{evans2021horizonstudycosmicexplorer, evans2023cosmicexplorersubmissionnsf} and the Einstein Telescope \citep[ET; ][]{Punturo2010ET}, should detect mergers at a rate $\sim 10^3$ times higher than that of current GW detectors at their so-called ``A+'' sensitivity. Hence, probing even the rarest systems should become possible \citep{gupta2024,evans2021horizonstudycosmicexplorer,evans2023cosmicexplorersubmissionnsf}.

\begin{figure} \label{fig:GW170817_comparison}
\includegraphics[width=0.5\textwidth]{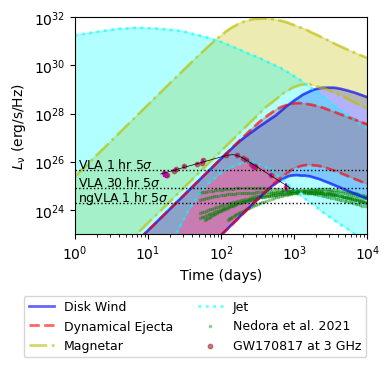}
\caption{
GW170817 observations at 3\,GHz (brown dots) and their phenomenological best fit from \citet{Makhathini_2021} (black solid line). The shaded areas enclose 90\% of the light curves simulated in this work for various ejecta components: the relativistic jet (cyan), the dynamical ejecta (red), the disk wind (blue), and the disk wind with magnetar injection (yellow). The horizontal lines show the sensitivity reached by the VLA and ngVLA with various integration times (1\,hr--30\,hr). In green we show the results of numerical simulations presented in \citet{Nedora_2021}. These fall outside of our 90\% containment regions for kilonova afterglows, but within our 99\% contours (not shown here). This is to be expected given that the
ISM density adopted by \citet{Nedora_2021,Nedora_2021b} for modeling GW170817 is on the low end of the range explored here \citep[and motivated by short GRB observations;][]{Fong2015}.}
\end{figure}

The fate of the remnant of GW170817 itself is still uncertain, though various lines of evidence suggest that a short-lived NS likely formed after the merger and then collapsed to a black hole \citep[BH;][]{Margalit_2017,Rezzolla_2018, Gill_2019,Bauswein_2017}. The field of GW170817 continues to be monitored in the radio band years after the merger, in search of late-time radio emission that could shed further light on the nature of its merger remnant via its imprints on the structure (energy and speed distribution) of the merger ejecta \citep{Troja_2020,Balasubramanian_2021,Nedora_2021,2025MNRAS.539.2654K}.
Indeed, 
broadly speaking, three main radio emission scenarios could be realized in a BNS merger. If the merger results in a prompt BH formation, we expect a relativistic jet to be launched, powering a non-thermal afterglow \citep[e.g.,][see also the cyan area in Figure~\ref{fig:GW170817_comparison}]{Sari_1998}. If a stable NS is formed in the  merger, it may power a relativistic jet \citep{2008MNRAS.385.1455M,2020ApJ...901L..37M} and/or a late-time radio flare associated with the fastest components of the ejected neutron-rich debris \citep[hereafter, kilonova ejecta; red and blue areas in Figure ~\ref{fig:GW170817_comparison};][]{Nakar_2011}. The flare brightness may be enhanced if the NS is sufficiently magnetized  \citep[yellow area in Figure~\ref{fig:GW170817_comparison};][]{Sarin_2022}. In between these two extremes is the case of a short-lived NS formed after the merger, which then collapses to a BH. In this case (which should be most similar to the case of GW170817; data points in Figure~\ref{fig:GW170817_comparison}), a radio afterglow from a relativistic jet (brown dots in Figure~\ref{fig:GW170817_comparison}) could be followed by a rather dim late-time radio flare. The last may be detectable once the emission from the jet fades, if the NS EoS is not too stiff \citep[green lines in Figure~\ref{fig:GW170817_comparison};][]{Nakar_2011,2015MNRAS.450.1430H,2016ApJ...819L..22H,2018ApJ...867...95H,2019MNRAS.485.4150B,Nedora_2021,Balasubramanian_2021,Balasubramanian_2022}.

In light of the above considerations, here we explore the detectability of radio counterparts of BNSs with total masses $M_{\rm tot}\lesssim 3$\,M$_{\odot}$, and 90\% GW localization area $\Omega \le 10$\,deg$^2$. While other works have analyzed prospects for detecting EM counterparts to BNS mergers \citep{han2025,Ronchini_2022,loffredo2024prospectsopticaldetectionsbinary,kaur2024detectingpromptafterglowjet,kunnumkai2025detectingelectromagneticcounterpartsligovirgokagra}, here we present a self-consistent approach (from GW detection to radio detection) that includes a variety of possible ejecta components (jets, dynamical ejecta, and disk winds) and focuses specifically on relatively well-localized systems where a short- or long-lived NS could be the product of the merger (as opposed to prompt BH formation scenarios). We stress that, by applying stricter requirements on the GW localization areas than most previous studies, our analysis emphasizes GW events for which: (i) the search for any optical counterpart could be done relatively easily \citep[i.e., with $\mathcal{O}(1)$ snapshots with large field-of-view facilities such as the Vera Rubin Observatory;][]{2019ApJ...873..111I,2024arXiv241104793A}; (ii) radio observations could be carried out (and discoveries enabled)  completely independently of an optical detection and localization. Hence, our analysis also quantifies the prospects for detecting radio counterparts beyond the limiting redshift of detectable kilonovae  \citep[estimated to be $z\lesssim 0.5$;][]{blum2022snowmass2021cosmicfrontierwhite,Andreoni_2022,loffredo2024prospectsopticaldetectionsbinary},  leveraging the enhanced survey speed of future radio arrays \citep[][]{Murphy_2017,selina2018generationlargearraytechnical,wilner2024keysciencegoalsgeneration,2019BAAS...51g.255H}.

Our paper is organized as follows. In Section~\ref{sec:style} we describe our methods. In Section~\ref{sec:results} we present our results. In Section~\ref{sec:conclusion} we summarize and conclude.

\section{Methods} \label{sec:style}
We generate a synthetic population of \gls{BNS}s with observationally-motivated properties, following a procedure similar to the one presented in \cite{gupta2024}. As described in more details in Section \ref{subsec:BNSMergerPopulation}, we restrict our analysis to BNSs with $M_{\rm tot}\le 3$\,M$_{\odot}$. We quantify the GW detectability and localization of these BNSs using \texttt{Gwbench} \citep{Borhanian_2021}. We also focus on BNSs that can be observed with a GW network signal-to-noise ratio (hereafter, $SNR$) $\ge 10$ and with 90\% confidence GW localization areas $\Omega \le 10$\,deg$^2$ (Section \ref{subsec:GWDetection}). 

For our study, we consider the GW detector networks listed in Table~\ref{tab:GWnetworks}. These include a network of LIGO-Hanford (H), LIGO-Livingston (L), and LIGO-India (A) \citep{UNNIKRISHNAN_2013,LSC_program_2023} at post-O5 sensitivities \citep{post-O5_report}. These are A\# sensitivities for LIGO-Hanford and Livingston, and A+ sensitivity \citep{2018LRR....21....3A} for LIGO India. Hereafter, we refer to this network as H-L-A. We consider this network for the specific purpose of comparing our estimates on future GW detector networks (see below) with a promising but realistic (for the next decade) current-generation network.  Next, we include networks comprised of CE with 20\,km (CE20) and 40\,km (CE40) arms, ET, and LIGO-India \citep{evans2023cosmicexplorersubmissionnsf,ngGWMPSACREport}. We do not consider all possible combinations of these detectors, but rather focus on networks (hereafter, CE40-ET-A; CE40-ET; CE40-CE20-A; CE40-CE20) recommended by the so-called ngGW Detector Concept Subcommittee of the National Science Foundation Mathematical and Physical Sciences Advisory Committee (MPSAC). This committee was tasked to assess and recommend a set of concepts for new GW observatories in the US \citep{ngGWMPSACREport}. Finally, we compare results for the MPSAC-recommended next-generation GW detector networks, with the network of three next-generation detectors considered in \citet{evans2023cosmicexplorersubmissionnsf} (hereafter, CE40-CE20-ET). The last represents one of the most ambitious networks (at least from the standpoint of US-based investments). For consistency with \cite{gupta2024} and \cite{ngGWMPSACREport}, we consider LIGO-India at A\# sensitivity every time that it is considered in a network with next-generation GW detectors like CE and ET.
  
Next, as described in detail in Section \ref{subsec:RadioCounterparts}, we estimate the detectability of potential radio counterparts associated with BNS ejecta using current and next generation US-based radio arrays operating at cm-wavelengths, such as the Karl G. Jansky Very Large Array \citep[VLA;][]{Thompson1980VLA,Perley_2011}, the next generation VLA \citep[ngVLA;][]{Murphy_2017,selina2018generationlargearraytechnical,wilner2024keysciencegoalsgeneration}, and the Deep Synoptic Array-2000 \citep[DSA-2000;][]{2019BAAS...51g.255H}. To this end, we assume that each BNS in our sample produces a successful jet with a structure (energy distribution as a function of polar angle from the jet axis) similar to that of GW170817 (Section \ref{sec:grbjet}). We simulate radio emission from jets using the code \texttt{Afterglowpy} \citep[][]{Ryan_2020}, with model parameter values motivated by observations of short GRB afterglows \citep{Fong2015}. We also use the fitting formulas provided by \cite{Coughlin_2019}  to predict, for each BNS, masses and speeds of the dynamical ejecta and disk winds. We then use the public code \texttt{Redback} to simulate potential radio counterparts from these ejecta components (Section \ref{sec:methods_modeling_the_kilonova_components}). Finally, we include a scenario where mergers can produce magnetized NS remnants (magnetars) that pump energy into the disk wind ejecta \citep[Section \ref{sec:Methods_MagnetarModels};][]{Sarin_2022}.

\begin{table*}
    \caption{GW detector networks considered in this study. The nominal location and orientation of all observatories are given in Table 2 of \cite{gupta2024} \citep[see also ][for an extensive discussion of detectors' location]{Daniel_2024} . These GW network configurations mirror those recommended by \cite{ngGWMPSACREport} and also discussed in \cite{evans2023cosmicexplorersubmissionnsf}. \label{tab:GWnetworks}}
    \begin{center}
\begin{tabular}{llc}
\hline
\hline
Network & Detectors & Notes\\
\hline
H-L-A & LHO(A\#) LLO(A\#) LIGO India(A+) & Post-O5 network \citep{post-O5_report} \\
CE40-ET-A & CE 40\,km ET LIGO-India(A\#) &Recommendation \#1 in \citet{ngGWMPSACREport}\\
CE40-ET & CE 40\,km ET & Recommendation \#2 in \citet{ngGWMPSACREport}\\
CE40-CE20-A & CE 40\,km CE 20\,km LIGO-India(A\#) & Recommendation \#3 in \citet{ngGWMPSACREport}\\
CE40-CE20 & CE 40\,km CE 20\,km & Recommendation \#4 in \citet{ngGWMPSACREport}\\
CE40-CE20-ET & CE 40\,km CE 20\,km ET & Three-ngGW-detector case in \citet{evans2023cosmicexplorersubmissionnsf}\\
\hline
\end{tabular}
\end{center}
\end{table*}

\begin{figure}
\includegraphics[width=\columnwidth]{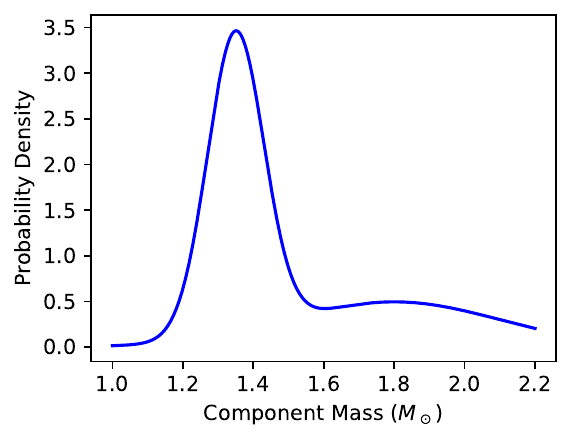}
\caption{Bi-modal Gaussian distribution of the NS component masses (see Equation~\ref{eq:prob_mass_distribution}). 
\label{fig:mass_prob_density}}
\end{figure}

\begin{figure}
\includegraphics[width=\columnwidth]{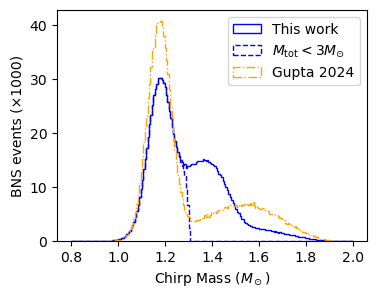}
\caption{A comparison of the (source-frame) BNS chirp-mass distributions in this work and that in \citet{gupta2024}. 
Here we focus on the BNS mergers which have the potential to form NS remnants (those with total masses $\le 3 M_{\odot}$, to the left of the dashed blue line). Hence, our BNS population is very similar to that of \cite{gupta2024}.
\label{fig:chirpmass_comparison}}
\end{figure}

\subsection{BNS Merger Population} \label{subsec:BNSMergerPopulation}
We start from the population of BNS mergers used in \citet{gupta2024}. The two NSs in each of the BNSs in this population are characterized by source-frame masses $m_1\ge m_2$. The corresponding total mass, mass ratio, and chirp mass of the BNS are defined as:
\begin{equation}
    M_{\rm tot} = m_2+m_1,
\end{equation}
\begin{equation}
    q = m_2/m_1
\end{equation}
and:
\begin{equation}
    \mathcal{M}=\frac{(m_1m_2)^{3/5}}{(m_1+m_2)^{1/5}},
\end{equation} respectively. 
The chirp mass and mass ratio distribution of the population used by \citet{gupta2024} is derived from a mass distribution of the NSs in the binary represented by a two-component Gaussian, truncated between 1\,M\,$_\odot$ and 2.2\,M$_\odot$:
\begin{equation}
p(m) = wN(\mu_L, \sigma_L) + (1 - w)N(\mu_R, \sigma_R), \label{eq:prob_mass_distribution}
\end{equation}
where $\mu_L = 1.35$\,M$_\odot$, $\sigma_L = 0.08$\,M$_\odot$, $\mu_R = 1.8$\,M$_\odot$, $\sigma_R = 0.3$\,M$_\odot$, and $w = 0.64$, consistent with the results of \cite{Farr_2020}. This distribution is plotted in Figure~\ref{fig:mass_prob_density}. In their work, \citet{gupta2024} impose that NSs from each of the two Gaussian components of the bi-modal distribution above only merge with other NSs from the same Gaussian component. Here, we relax this assumption by re-drawing NS component masses independently of each other. Overall, this results in a different chirp mass and mass ratio distribution of our BNSs (Figure~\ref{fig:chirpmass_comparison}), while the distribution of the other parameters (discussed below) mirrors that in \citet{gupta2024}. 

The dimensionless NS spins are defined as:
\begin{equation}
    \chi_i=\frac{cI_i\omega_i}{Gm_i^2},
\end{equation}
where $I$ is the moment of inertia, $\omega$ is the NS angular spin frequency, $m$ is the NS mass, $G$ is the gravitational constant, and $c$ is the speed of light. The index $i$ is used to refer to the component NSs. The dimensionless spins are uniformly distributed from 0 to 0.1 \citep{Zhu_2018}. 

Each BNS is also assigned a redshift $z$, sky position (Right Ascension, Declination), and inclination angle of the binary axis relative to the line of sight to Earth. The redshift distribution of the BNSs is determined by the Madau-Dickinson star-formation rate \citep{Madau_2014,Talbot_2019}, given by Equation 1 of \cite{gupta2024}. For consistency with \citet{gupta2024}, the local merger rate is assumed to be equal to $320$\,Gpc$^{-3}$ yr$^{-1}$ \citep{Beniamini_2021,Saleem_2017,Abbott_2017_GW170817}, with the merger rate as a function of redshift  proportional to the local merger rate. Rates as a function of redshift scale linearly with the local merger rate and can be adjusted to reflect the most recent estimates of 10-1700 Gpc$^{-3}$yr$^{-1}$ \citep{LVKObservingScenarios2020,2023PhRvX..13a1048A,loffredo2024prospectsopticaldetectionsbinary}. 

\begin{deluxetable*}{ccc}
\tablecaption{A summary of the parameters characterizing our BNS population. \label{tab:BNS_population}}
\tablecolumns{3}
\tablewidth{0pt} 
\tablehead{
\colhead{Symbol} & \colhead{Description}& \colhead{Modeled Value}
}
\startdata
$\theta_{\rm obs}$=min($\iota,\pi-\iota)$ & Viewing angle & $\cos (\theta_{\rm obs})$ Uniformly distributed in [-1, 1]\\
$z$ & Redshift & Madau-Dickinson S.F.R.\\
$d_L$ & Luminosity distance & corresponding to z, assuming $H_0=67.66~\text{km}/\text{s}/\text{Mpc}$\\
$m_{1,2}$ & Component masses & Equation~\ref{eq:prob_mass_distribution}\\
RA & Right Ascension & Uniformly distributed in [0, $2\pi$] \\
Dec & Declination & $\cos({\rm Dec})$ Uniformly distributed in [-1, 1]\\
$\chi_{1,2}$ & Component spins & Uniformly distributed in [0, 0.1] \\
$\psi_{1,2}$ & Polarization angle & Uniformly distributed in [0, $2\pi$] \\
$p_0$ & Initial NS rotational period & Uniformly distributed in $\left[5\times 10^{-4}, 1\times 10^{-3}\right]$\,s\\
$\chi_0$ & Initial angle between magnetic and rotation axes & $\frac{\pi}{2}$ \\
$R_1$ & Radius of the first component NS & Uniformly distributed in $\left[10, 15\right]$ km\\
$R_2$ & Radius of the second component NS & Uniformly distributed in $\left[10, 15\right]$ km\\
\enddata
\end{deluxetable*}

\begin{figure}
\includegraphics[width=\columnwidth]{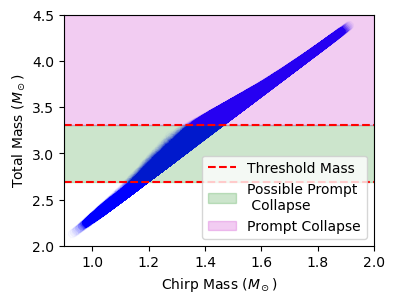}
\caption{A scatter plot of the chirp mass versus the total mass for all of the \gls{BNS}s in our in 1-yr simulation (blue). The minimum and maximum values of the threshold total mass $M_{\rm thr}$ for prompt collapse is indicated in red.  The mass region corresponding to prompt collapse, possible prompt collapse, and a NS remnant are indicated in magenta, green and white, respectively. Hereafter, we assume $M_{\rm thr}\approx 3\,M_\odot$ and consider BNSs with total masses below this threshold value to be candidates for leaving behind NSs as their merger remnants. Below $2.17\,M_{\odot}$, a long-lived NS could form.}
\label{fig:collapse_percentages}
\end{figure}

The properties of our simulated BNS mergers are summarized in Table~\ref{tab:BNS_population}. 
Our 1-yr BNS merger population contains $\approx 1.3\times10^6$ BNSs merging in 1\,year across all redshifts and sky positions. 
It is central to this study to examine the multi-messenger detectability of all systems that could result in a NS remnant i.e., systems that do not undergo prompt BH collapse. The condition for prompt collapse is determined by numerical simulations to be when the total mass of the system, $M_{\text{tot}}$ exceeds a threshold mass $M_{\text{thr}}$ given by the TOV limit \citep[][]{Margalit_2019, Bauswein_2013,Bauswein_2017}:
\begin{equation}
    M_{\text{tot}}> M_{\text{thr}} = \left( 2.38-3.606\frac{GM_{\text{TOV}}}{c^2R_{1.6}}\right)M_{\text{TOV}},\label{eq:mthr}
\end{equation}
where $R_{1.6}$ is the radius of a $1.6M_\odot$ NS (which depends on the assumed EoS). We calculate $M_{\text{\rm thr}}$ by considering the minimum and maximum possible values of $M_{\rm TOV}$ and $R_{1.6}$ \citep{De_2018,Margalit_2017,Antoniadis_2013,Bauswein_2017}:
\begin{align}
    2.01\,M_\odot \le& M_{\rm TOV}\le2.17\,M_\odot\\
    10.3\,\text{km} \le&R_{1.6}\le13.5\,\text{km},
\end{align}
which imply:
\begin{equation}
2.69\,M_\odot \leq M_{\text{thr}} \leq 3.31\,M_\odot. 
\end{equation}
We plot the range for $M_{\text{thr}}$ in horizontal red lines in Figure~\ref{fig:collapse_percentages}, together with a scatter plot of the chirp masses and total masses of all of the \gls{BNS}s in our 1-yr simulation. This is similar to and consistent with Figure 4 of \cite{Margalit_2019}. Given the uncertainty on $M_{\rm TOV}$ and $R_{1.6}$, we indicate the regions in which prompt collapse is inevitable (magenta), a NS remnant is inevitable (white), and the total mass region where either outcome could occur (green). Of the BNS mergers in our 1-yr population, we find that 23\%--80\% would form a NS remnant, and 20\%--77\% would experience prompt collapse.

\begin{table*}
    \caption{Properties of the BNS populations used in this study. For all of the populations, we assume a local BNS merger rate of 320\,Gpc$^{-3}$\,yr$^{-1}$. From left to right, we report: the total number of BNSs in the simulated population; the median redshift of the population; the minimum redshift; the number of BNSs in the population  with $M_{\rm tot} \le 3\,M_{\odot}$ (these are candidates for NS post-merger remnants); the number of BNSs in the population  with $M_{\rm tot} \le 2.17\,M_{\odot}$ (these are candidates for long-lived NS post-merger remnants); the number of BNSs in the population with $0.7 \le q \le 1$. We note  that the 50-yr simulation was run specifically to better sample the low-$z$ events, so it only covered redshifts $z\le  0.13$. As a result, the total number of BNS mergers (first column) for this population is much smaller than that of the 1-yr population. The 700-yr population was simulated specifically to better sample the low-mass end of the BNS merger distribution. Hence, by construction, this population only extends to $M_{\rm tot}\le 2.17\,M_{\odot}$ and contains a total number of events (first column) of $\approx 700\times 23$ i.e., $\approx 700\times$ the number of BNSs with $M_{\rm tot}\le 2.17\,M_{\odot}$ in the 1-yr population. \label{tab:original_stat}}
    \begin{center}
\begin{tabular}{ccccccc}
\hline
\hline
Total & $z_{50\%}$  & $z_{\rm min}$ & $z\le 0.13$ & $M_{\rm tot}\le 3\,M_{\odot}$ & $M_{\rm tot}\le 2.17\,M_{\odot}$ & $0.7\le q \le 1$  \\
\hline
\multicolumn{7}{c}{\bf 1-yr population}\\
$1.3\times 10^6$ & 2.1 & $4.3\times 10^{-2}$ & $1.7\times 10^5$ (13\%) & $7.4\times 10^5$ (58\%) & $2.3\times 10^1$ (0.0018\%) & $7.3\times10^5$ (57\%)\\
\hline
\multicolumn{7}{c}{\bf 50-yr population}\\
$1.4\times 10^4$ & 0.10 & $5.6\times10^{-4}$ & $1.4\times 10^4$ (100\%) & $7.9\times 10^3$ (58\%) & $0$   & $7.7\times 10^3$ (57\%) \\
\hline
\multicolumn{7}{c}{\bf 700-yr population}\\
$1.6\times 10^4$ & 2.1 & $7.6\times 10^{-2}$ & $2.2\times 10^3$ (13\%) & $1.6\times 10^4$ (100\%)& $1.6\times 10^4$ (100\%) & $1.6\times 10^4$ (100\%) \\
\hline
\hline
\end{tabular}
\end{center}
\end{table*}

In order to get better statistics at lower values of $M_{\rm tot}$ and at lower redshifts, we also supplement the 1-year population with multi-year simulations of only the relevant portion of parameter space. More specifically, for low-mass BNSs \citep[$M_{\rm tot} \le 2.17\,M_{\odot}$, chosen for consistency with the upper limits of $M_{\rm TOV}$;][]{De_2018,Margalit_2017,Antoniadis_2013,Bauswein_2017}, we simulate a 700-year population of BNS mergers. This yields a total of $\approx 1.6\times10^4$ BNSs in this portion of the parameter space, for which we extract an average annual rate when merging results with the broader 1-yr BNS population. For investigations into BNS mergers at $z\le 0.13$, we run a 50-year simulation (this yields a total of $\approx 1.4\times10^4$ BNSs in this portion of the parameter space), and take 50-year averages on the annual rates. These specific populations are independent of the population of our 1-year sample, but are generated consistently with it in terms of redshift and component mass distributions. We provide a summary of our populations in Table \ref{tab:original_stat}.

\subsection{GW Detection and Localization} \label{subsec:GWDetection}
\begin{figure}
\includegraphics[width=\columnwidth]{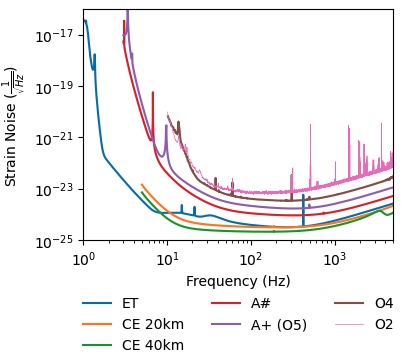}
\caption{Sensitivity curves for the detectors used in our study, taken from \texttt{Gwbench} \citet{Borhanian_2021} and consistent with \citet{ngGWMPSACREport} for next-generation detectors, compared with the sensitivity of the LIGO detectors during their second observing run O2 \citep[the run that brought to the discovery of GW170817;][]{Kissel_O2_sens,2021SoftX..1300658A} and to their nominal sensitivities during the most recent observing run O4 \citep{LVKObservingScenarios2020,LVK_observing_scenario_paper}. 
\label{fig:DetectorSensitivities}}
\end{figure}

To determine the potential for GW detection and localization of the BNSs in our population, we use \texttt{Gwbench} \citep{Borhanian_2021}, a software package that uses the Fisher information formalism as a fast tool for parameter estimation. 
More specifically, for each detector network, we use \texttt{Gwbench} to derive, for each BNS merger in our population: the 90\% confidence sky localization $\Omega$, the network $SNR$, the observing angle $\theta_{\rm obs}=min(\iota,\pi-\iota)$ \citep{2018ApJ...860L...2F} where $\iota$ is the inclination angle (i.e., the angle between the direction perpendicular to the plane of the BNS orbit and the line of sight to Earth), and the luminosity distance $d_L$. In the formalism used by \texttt{Gwbench}, the signal-to-noise ratio measured by the ith-detector ($SNR_{i}$) is defined by Equation 3.6 in \citet{Borhanian_2021} as:
\begin{equation}
SNR^2_i = 4\int_{\rm f_{min,i}}^{\rm f_{max}} df \frac{|H_i(f)|^2}{S_{n,i}(f)},
\end{equation}
where $H_i$ is the GW strain in the frequency-domain measured by the i-th detector (combining the waveform polarizations with the detector’s antenna patterns and location phase factor), and $S_{n,i}$ is the power spectral density of the ith-detector. Also in the above Equation: ${\rm f_{min,i}}$ is the low-frequency cutoff for the $i$-detector, here assumed to be 1\,Hz for ET and 5\,Hz for CE (Figure \ref{fig:DetectorSensitivities}); ${\rm f_{max}}$ is the frequency of the innermost stable circular orbit. The $SNR$ of a network of GW detectors is given by Equation 3.12 in \citet{Borhanian_2021} as:
\begin{equation}
SNR^2 = \sum_{i=1}^{N_d}SNR^2_{i},
\end{equation}
where $N_d$ is the number of detectors in the network. We consider a merger to be detected when $SNR \ge 10$. For each detector we use the sensitivity curve provided in \texttt{Gwbench} (Figure~\ref{fig:DetectorSensitivities}). Given that the locations of next generation observatories are still undecided, we follow the prescriptions given in Table 2 of \cite{gupta2024}.

We note that a recent study comparing \texttt{Gwbench} (which is based on the Fisher
matrix---a quadratic approximation reliable in the strong signal limit) and other Bayesian parameter estimation methods  has demonstrated strong
consistency in $SNR$ and sky localization area estimations \citep{2025ApJ...985L..17P}. Hence, we consider the \texttt{Gwbench} results accurate enough for our study.

\begin{deluxetable}{ccccc}
\tablecaption{The rms sensitivity of the VLA, ngVLA and DSA-2000 \citep{DSA_overview, hallinan2019dsa2000radiosurvey} at different frequencies. For the VLA, we use the nominal sensitivity for a 1\,hr-long observation (including typical overhead) in A configuration. For the ngVLA, we use the naturally weighted 1\,hr sensitivity values reported in \cite{ngVLA_sens} which assume pointed observations. We report in parenthesis the sensitivity that can be reached with the ngVLA in survey mode over a 10\,deg$^2$ area in about 10\,hrs \citep{2018ASPC..517...15S,wilner2024keysciencegoalsgeneration}. We do not account for potential bandwidth reduction due to radio frequency interference.  \label{tab:ngVLAsensitivity}}
\tablecolumns{4}
\tablewidth{\columnwidth}
\tablehead{
\colhead{\makecell{Frequency \\ (GHz)}} &  \colhead{\makecell{VLA \\ ($\mu$Jy)}} & 
\colhead{\makecell{ngVLA \\ ($\mu$Jy)}} &\colhead{\makecell{DSA-2000 \\ ($\mu$Jy)}} 
}
\startdata
1.4 & 10 & -  & 1 & \\
2.4 (2.2) & - &0.24 (1.0) &  - \\
3.0 & 5.0 & -  & - \\
8.0 & 3.6 & 0.14  & - \\
16 & 4.9 &0.16 & - \\
22 & 8.4 & - & - \\
27 & - &0.17 (10) & - \\
33 & 8.5 &- & - \\
41 & - &0.21  & - \\
45 & 23 &- & - \\
93 & - &0.40  & - \\
\enddata
\end{deluxetable}

\subsection{Radio Counterpart Population and Detection } \label{subsec:RadioCounterparts}
In modeling BNS radio counterparts, we consider only BNSs with $M_{\rm tot} \leq 3\,M_\odot$ (i.e., systems that at least in principle could leave behind short- or long-lived NSs as their merger remnants). The number of BNSs in our original samples that meet this requirement are 58\% of the $1.3\times10^6$ events in the 1-year sample, 58\% and 100\% of the $1.4\times10^4$ and $1.6\times10^4$ events in the 50-yr and 700-yr sample, respectively (see Table \ref{tab:original_stat}).
Only for assessing the detectability of kilonova (disk wind and dynamical ejecta) radio afterglows, we further restrict our analysis to BNSs with $0.7 \le q \le 1$ for which the approximations described in Section \ref{sec:methods_modeling_the_kilonova_components} are applicable \citep{Coughlin_2019, Margalit_2019}. When considering the cuts placed on both $M_{\rm tot}$ and $q$, we are left with $\approx 7.3\times10^5$ BNSs in a 1-yr sample i.e., 57\% of the 1-yr population (see Table \ref{tab:original_stat}). We then evaluate GW detectability and localization accuracy as described in Section \ref{subsec:GWDetection} and as a function of the chosen detector network (Table \ref{tab:GWnetworks}). We consider worthy of EM follow up only BNSs with network $SNR\ge 10$ and sky localization regions are $\Omega \le 10$\,deg$^2$. This GW localization requirement is stricter than what was considered in previous studies \citep[e.g.,][]{kaur2024detectingpromptafterglowjet}, and would enable optical follow up with $\mathcal{O}(1)$ pointing(s) from a facility with a $\approx 10$\,deg$^2$ field-of-view like Rubin \citep{Daukantas2024}. At the same time, because an area of $\approx 10$\,deg$^2$ can be covered relatively quickly ($\lesssim 10$\,hr) with $\mu$Jy-level sensitivity by future radio arrays of enhanced survey speed such as the ngVLA \citep[see Table \ref{tab:ngVLAsensitivity};][]{Murphy_2017,selina2018generationlargearraytechnical,wilner2024keysciencegoalsgeneration} and/or the DSA-2000 \citep[][]{2019BAAS...51g.255H}, our analysis also provides realistic estimates for GW plus radio detections that are independent of localizations at other EM wavelengths. 

\begin{figure}
\includegraphics[width=\columnwidth]{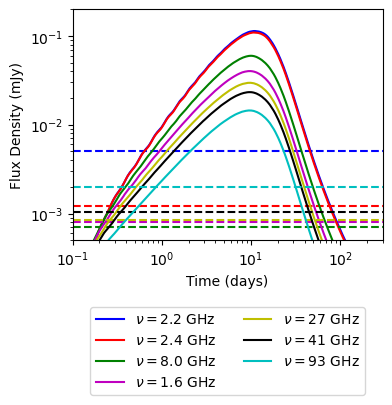}
\caption{An example light curve generated using \texttt{afterglowpy} at ngVLA frequencies (see Table \ref{tab:ngVLAsensitivity}). The threshold flux densities at which detections can be made at the $5\sigma$ confidence level are plotted with dashed lines. 
For this example we assume: $E_0 = 1.8 \times 10^{52}\,\mathrm{erg}$, $n_{\rm ISM} = 9.5 \times 10^{-1}\,\mathrm{cm}^{-3}$, $z = 0.4$, $\theta_{\rm obs} = 0.36$\,rad, $\theta_{c} = 0.06$\,rad, $\epsilon_e = 1.9\times 10 ^{-2}$, $\epsilon_B=8.7\times 10^{-2}$, $d_L=7.1\times 10^{27}$ cm, $p=2.16$, and $\theta_w=2.4\times10^{-1}$\,rad. We neglect the effects of synchrotron self-absorption. }
\label{fig:Flux_density_v_time}
\end{figure}

As described in detail in Sections \ref{sec:grbjet}-\ref{sec:Methods_MagnetarModels} we evaluate the radio detectability BNSs that are worthy of EM follow-up and observable with US-based current and/or future radio facilities \citep[sky location with declination $\ge-40$\,deg][]{VLASS_NRAO,Byrne_2024} by comparing the predicted radio flux density at a specific frequency to a threshold taken to be $5\times$ the estimated root-mean-square (rms) sensitivity at that frequency (Table~\ref{tab:ngVLAsensitivity}). If the predicted flux density falls above this threshold at any time since merger, the radio counterpart is considered detected (Figure~\ref{fig:Flux_density_v_time}).  

\subsection{GRB Jet Afterglows}
\label{sec:grbjet}
We work under the assumption that all BNSs are capable of launching relativistic jets. This is motivated by the association of GW170817 with a GRB, and by the fact that BNS and (short) GRB rates are compatible with each other within current uncertainties \citep{2022LRR....25....1M}. In the future, our results could be rescaled for a jet formation efficiency factor, once such factor becomes better constrained observationally. In fact, systematic radio observations of a large sample of GW-detected BNSs are expected to provide a geometry-independent constraint on the fraction of BNSs that produce successful relativistic jets (Section~\ref{subsec:Jet_results}). 

To model the jet radio afterglows, we use the redshifts $z$ of the GW-detected BNSs in our population (Section \ref{subsec:BNSMergerPopulation}), and their viewing angles $\theta_{\rm obs}=min(\iota,\pi-\iota)$, where $\iota$ is the BNS inclination angle (Section \ref{subsec:BNSMergerPopulation}). 
The corresponding  afterglow light curves are generated using \texttt{afterglowpy} with input parameters as specified in Table~\ref{tab:Afterglowpyinputs}. We assume Gaussian jets with energy profiles:
\begin{equation}
E(\theta)_{\text{Gaussian}} = E_0 \exp\left(-\frac{\theta^2}{2 \theta_c^2}\right),
\end{equation}
where $\theta$ is the angle from the jet axis, $\theta_c$ and \(E_0\) are the jet core opening angle and isotropic equivalent energy, respectively.  

Because \texttt{afterglowpy} does not include effects related to synchrotron self-absorption (SSA), which can suppress the observed radio flux at the lower radio frequencies, here we adopt a simplified approach to account for SSA. Namely, we calculate $\nu_{\rm SSA}$ following simple analytical prescriptions for on-axis relativistic jets:
\begin{equation}
\nu_{\rm SSA}\propto \epsilon_e^{-1} \epsilon_B^{1/5}E_0^{1/5}n_{\rm ISM}^{3/5}.
\label{eq:nussa}
\end{equation}
In the above Equation, $\epsilon_B$ and $\epsilon_e$ are the fraction of energy going into magnetic fields and accelerated electrons, respectively, and $n_{\rm ISM}$ is the density of the interstellar medium. The normalization constant for the SSA frequency above varies by almost an order of magnitude across the literature (compare e.g. Equation (22) in \citet{1999ApJ...527..236G} with Equation (60) in \citet{2000ApJ...543...66P}). Hence, here we first compute relativistic jet radio light curves neglecting SSA effects, and then explicitly discuss the potential impact of SSA on these light curves using the range of values for $\nu_{\rm SSA}$ implied by different normalizations for Equation \ref{eq:nussa}. We note that this approach neglects corrections for viewing angle and jet structure effects \citep[for a discussion of these effects, see e.g.][]{2024ApJ...977..181S}.

\begin{deluxetable*}{ccc}
\tablecaption{Input parameter values for \texttt{afterglowpy}. These parameters are motivated by GW170817 observations \citep{Makhathini_2021,Troja_2017,Troja_2019,Troja_2020}, and by the constraints set by the analysis of 103 other GRBs \citep{Fong2015}. See text for discussion. \label{tab:Afterglowpyinputs}}
\tablecolumns{3}
\tablewidth{0pt}
\tablehead{
\colhead{Symbol} & \colhead{Description}& \colhead{Modeled Value}
}
\startdata
jetType & $\theta$-dependent jet structure & Gaussian \\
specType & Synchrotron spectrum & 0 \\
$\theta_{\rm obs}$ & Viewing angle & See Table \ref{tab:BNS_population}\\
$z$ & Redshift & See Table \ref{tab:BNS_population}\\
$E_0$ & Isotropic-equivalent energy & Uniformly Distributed $[4\times 10^{49}, 3.4\times 10^{52}]$\,erg \\
$\theta_c$ & Jet core half-opening angle & Gaussianly Distributed $\mu = 8^\circ$, $\sigma = 5^\circ$ \\
$n_{\rm ISM}$ & ISM density & Uniformly Distributed [$3\times 10^{-3}$, 1] $\text{cm}^{-3}$ \\
$p$ & Power-law index of the electron energy distribution & Uniformly Distributed [2.15,2.79] \\
$\epsilon_e$ & Fraction of energy given to electrons & Uniformly Distributed [0.01,0.1]\\
$\epsilon_B$ & Fraction of energy given to magnetic field & Uniformly Distributed [0.01,0.1] \\
$\xi_N$ &Fraction of accelerated electrons  & 1
\enddata
\end{deluxetable*}

\begin{deluxetable*}{ccc}
\tablecaption{The \texttt{Redback} functions that we utilize for computing the radio afterglow of each kilonova ejecta component (dynamical and disk wind), and their input parameters. The magnetar luminosity $L$ is the output of the \texttt{magnetar\_luminosity\_evolution} function of the \texttt{magnetar\_models}. It is one of the input parameters for the \texttt{\_ejecta\_dynamics\_and\_interaction} function of the \texttt{magnetar\_driven\_ejecta\_models} which is used to calculate the maximum kinetic energy of the ejecta $E_k$. The last is an input for the  \texttt{kilonova\_afterglow\_redback} function, which is used to generate the kilonova radio light curves.  \label{tab:redbackinputs}}
\tablecolumns{3}
\tablewidth{0pt}
\tablehead{
\colhead{Symbol} & \colhead{Description}& \colhead{Modeled Value}
}
\startdata
\hline
\multicolumn{3}{c}{\textbf{ magnetar\_luminosity\_evolution()}} \\
\hline
$B_{int}$ & Internal magnetic field & Uniformly Distributed $\left[B_{ext}, \frac{1}{\sqrt{3}} \times 10^{17}\right]$ G\\
$B_{ext}$ & External dipole magnetic field & Uniformly Distributed $\left[10^{14}, 10^{16}\right]$ G\\
$p_0$ & Initial rotational period & Uniformly Distributed $\left[5\times 10^{-4}, 1\times 10^{-3}\right]$ s\\
$\chi_0$ & Initial inclination angle & $\frac{\pi}{2}$ rad\\
$R_r$ & Radius of the remnant & Uniformly Distributed $\left[10, 15\right]$ km\\
$I$ & Moment of inertia & See Equation~\ref{Eq:MOI}\\
$\kappa_\gamma$ & Gamma-ray opacity for leakage efficiency ($\mathrm{cm}^2\,\mathrm{g}^{-1}$) & 0.1\\
\hline
\multicolumn{3}{c}{\textbf{ \_ejecta\_dynamics\_and\_interaction()}} \\
\hline
$M_{ej}$ & Ejecta Mass (Dynamical or Disk Wind) &See Equations~\ref{eq:disk_mass} and ~\ref{Eq:dyn_ejecta_mass}\\
$V_{ej}$ & Initial Ejecta Velocity & See Equation~\ref{Eq:ejeca_velocity}, or $\left[0.15, 0.2\right]$ c\\
$R_{ej}$ & Ejecta radius & $10^{9}$ cm\\
$\kappa$ & Opacity ($\mathrm{cm}^2\,\mathrm{g}^{-1}$) & 1\\
$n_{\rm ISM}$ & ISM density & See Table \ref{tab:Afterglowpyinputs}\\
$L$ & Magnetar Luminosity & From magnetar\_luminosity\_evolution() \\
pair\_cascade\_fraction & & 0.1\\
pair\_cascade\_switch & & True\\
use\_gamma\_ray\_opacity & & True\\
$\kappa_\gamma$ & Gamma-ray opacity for leakage efficiency ($\mathrm{cm}^2\,\mathrm{g}^{-1}$) & 0.1\\
\hline
\multicolumn{3}{c}{\textbf{ \_kilonova\_afterglow\_redback()}} \\
\hline
$z$ & redshift & See Table \ref{tab:BNS_population}\\
$E_k$ & Maximum kinetic energy & From \_ejecta\_dynamics\_and\_interaction()\\
$M_{\rm dyn}$ & Dynamical ejecta mass & See Equation~\ref{Eq:dyn_ejecta_mass}\\
$0.5\times M_{\rm disk}$  & Disk wind mass & See Equation~\ref{eq:disk_mass}\\
$n_{\rm ISM}$ & ISM density & See Table \ref{tab:Afterglowpyinputs} \\
$\epsilon_e$ & Fraction of energy given to electrons & See Table \ref{tab:Afterglowpyinputs}\\
$\epsilon_B$ & Fraction of energy given to magnetic field & See Table \ref{tab:Afterglowpyinputs} \\
$p$ & Electron energy distribution index & See Table \ref{tab:Afterglowpyinputs}
\enddata
\end{deluxetable*}

\subsection{Kilonova Afterglows}
\label{sec:methods_modeling_the_kilonova_components}
Neutron rich merger ejecta undergo r-process nucleosynthesis, and radioactive decay of the ashes of the r-process is thought to  power kilonovae, UV/optical/infrared transients lasting hours
to weeks \citep[e.g., ][and references therein]{2019LRR....23....1M}. This same ejecta can have so-called fast tails that can power non-thermal radio afterglows and hence can be a source of radio emission \citep{Nakar_2011,2015MNRAS.450.1430H,2016ApJ...819L..22H,2018ApJ...867...95H,2019MNRAS.485.4150B,Nedora_2021}. 

The neutron-rich debris of BNS mergers is typically modeled as a two-component ejecta: the dynamical (red kilonova) component made of material ejected dynamically during the merger via tidal torques and shock-driven mechanisms; and the disk-wind (blue kilonova) component launched on a viscous timescale by the remnant accretion disk, driven by neutrino emission, magnetic turbulence, and viscous processes \citep[e.g.,][and references therein]{Metzger_2014,2015MNRAS.449..390F,2024PhRvD.110b4003F}. 
We use the analytical approximation presented in \citet{Coughlin_2019} to estimate the ejecta mass and the initial ejecta velocity of these two components. Following Equation 1 in \citet{Coughlin_2019}, the disk mass is given by:
\begin{eqnarray}
\label{eq:disk_mass}
\log_{10}(M_{\rm disk}) 
= ~~~~~~~~~~~~& \\max\left[ -3, a\left(1 + b\tanh\left(\frac{c-M_{\rm tot}/M_{\rm thr}}{d}\right) \right) \right], &
\end{eqnarray}
where $a=-31.335$, $b=-0.976$, $c=1.0474$, and $d=0.05957$. In the above Equation, $M_{\text{tot}}$ is the total mass of the BNS, and $M_\text{thr}$ is the threshold for prompt collapse into a BH \citep[see Equation \ref{eq:mthr};][]{Bauswein_2013,Bauswein_2017}.  
Following \citet{Margalit_2019}, hereafter we set $M_{\text{TOV}}=2.1M_\odot$, and we assume that $50\%$ of the disk mass is accelerated away in the disk wind:
\begin{equation}
M_{\rm wind} \approx 0.5\times M_{\rm disk}.
\label{eq:mwind}
\end{equation}
The disk wind velocity values are not well constrained so, following \citet{Sarin_2022}, we draw them from a uniform distribution in the range $0.15c$ to $0.2c$. 

\begin{figure}
    \centering
    \vbox{
        \includegraphics[width=\columnwidth]{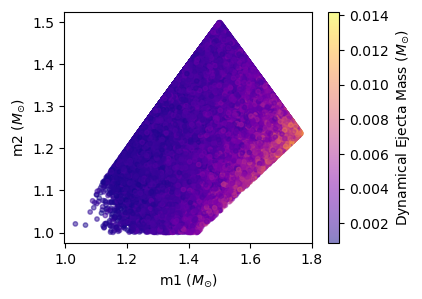}
        \includegraphics[width=\columnwidth]{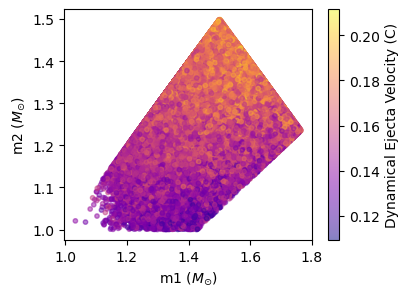}   
    }
    \caption{ Top: The dynamical ejecta mass, calculated from Equation~\ref{Eq:dyn_ejecta_mass}. Bottom: The dynamical ejecta velocity as calculated from Equation~\ref{Eq:ejeca_velocity}. We only show the region of parameter space in which $0.7\leq q\leq 1$ and $M_{\text{tot}} \leq 3 M_\odot$. }
    \label{fig:dynamical_velocity_and_mass}
\end{figure}
\begin{figure}
\includegraphics[width=\columnwidth]{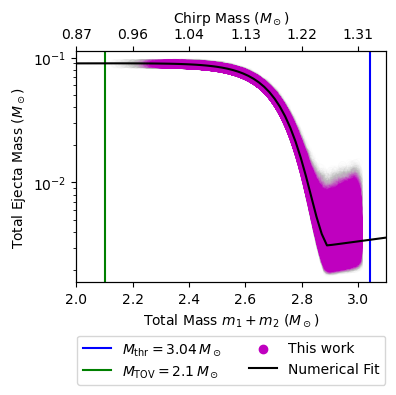}
\caption{Total ejecta mass (dynamical ejecta plus disk wind; purple) as a function of the total mass $M_{\rm tot}$ for BNSs with $0.7\leq q \leq 1$ \citep{Margalit_2019}, and $M_{\text{tot}}\le3M_{\odot}$. The top horizontal axis shows the chirp mass for $q=1$. In black, we plot the analytical fit given by Equations~\ref{eq:disk_mass} and~\ref{Eq:dyn_ejecta_mass}, assuming the component NSs each have a radius of $12.2$\,km \citep{Coughlin_2019}. We mark the $M_{\rm TOV}$ (green line) and $M_{\rm thr}$ (blue line) for a $1.6\,M_{\odot}$ NS with a 12\,km  radius \citep{Margalit_2019}. 
\label{fig:Metzger-like_ejecta_mass}}
\end{figure}

The dynamical ejecta mass can be estimated using Equation 2  in \citet{Coughlin_2019}:
\begin{equation}\label{Eq:dyn_ejecta_mass}
    log_{10}(M_{\rm dyn}) = \left[a\frac{(1-2C_1)m_1}{C_1} + b m_2\frac{m_1}{m_2}^n +\frac{d}{2}\right] + \left[1 \leftrightarrow 2\right],
\end{equation}
where $a=-0.0719$, $b=0.2116$, $d=-2.42$, $n=-2.905$. $m_{1,2}$ and $C_{1,2}$ represent the masses and compactness of the component NSs respectively, and $[1\leftrightarrow 2]$ indicates inter-changeability of stars. The compactness is defined as:
\begin{equation}
C_i = \frac{G m_i }{R_i c^2}.
\end{equation}
We note that the radius $R_i$ of the component NSs is needed to find the compactness values $C_1$ and $C_2$. Hence, we draw values of the NS radii from a uniform distribution between 10\,km and 15\,km. This way, while we remain agnostic to the details of any individual \gls{EoS}s, all of the BNSs in our sample can be reasonably described by the range of \gls{EoS}s shown in Figure 1 of \citet{Biswas_2022}. The properties and detectability of the dynamical ejecta as it relates to specific EoSs are explored in greater detail in \citet{rosswog2025fastdynamicejectaneutron}. The dependence of the ejecta mass on the component masses is shown in the top panel of Figure~\ref{fig:dynamical_velocity_and_mass}. The dynamical ejecta mass is greater in instances of lower mass-ratios.

In the case of the dynamical ejecta, we calculate the initial ejecta velocity as:
\begin{equation} \label{Eq:ejeca_velocity}
    v = \left[ a(1+cC_1)\frac{m_1}{m_2} + \frac{b}{2} \right] + \left[1 \leftrightarrow 2\right],
\end{equation}
where $a=-0.309$, $b=0.657$, $c=-1.879$ \citep{Coughlin_2019}.
 The dependence of the initial ejecta velocity is shown in the bottom panel of Figure~\ref{fig:dynamical_velocity_and_mass}.  The initial dynamical ejecta velocity increases in cases of higher total mass, and higher mass-ratios.

In Figure~\ref{fig:Metzger-like_ejecta_mass}, we plot the total ejecta mass:
\begin{equation} \label{Eq:total_ejecta_mass}
M_{\text{ejecta}} = M_{\text{dyn}} + 0.5\times M_{\text{disk}},
\end{equation}
for every BNS in our simulation for which the kilonova can be reasonably modeled. As evident from this Figure, the total ejecta mass is dominated in by the disk wind at smaller values of $M_{\rm tot}$ and by the dynamical ejecta mass at larger values of $M_{\rm tot}$. 

We use the \texttt{Redback}  software package \citep{sarin2024redback} to generate radio light curves for the disk wind and the dynamical ejecta components. For each BNS merger, we assume that both components are present and non-interacting, and we measure their detectability independently. The \texttt{Redback} input parameters are taken from our GW simulation when possible, and otherwise motivated by both GW170817 and the observed short GRB population \citep[see the last column of Table \ref{tab:redbackinputs} for details]{Fong2015}. For each BNS, model parameters common to both the kilonova and the jet models are kept consistent (see Table~\ref{tab:Afterglowpyinputs} and Table~\ref{tab:redbackinputs}). As in the case of the jet, we consider the kilonova to be detectable at all times in which the light curve falls above the sensitivity for a $5\times$\,rms detection for the ngVLA, given in Table~\ref{tab:ngVLAsensitivity}. Details on the use of the \texttt{Redback} software package are available in Appendix~\ref{sec:Appendix} and Figure \ref{fig:Figure8}, where we discuss a basic consistency check done to ensure agreement between this study and the results by \citet{Sarin_2022}.

The kilonova afterglow computation in \texttt{Redback} depends only on the ejecta mass, and the maximum kinetic energy of the ejecta (so, technically speaking, the code is blind to whether a given component is dynamical ejecta or disk wind). In the absence of energy injection from a long-lived merger remnant (magnetar), the maximum velocity reached by the ejecta is close to the initial velocity and never reaches the ultra-relativistic regime \citep[see e.g. Equation 6 in ][where $L_{\rm EM}$ is negligible or null in the no magnetar case]{Sarin_2022}. 

\begin{figure*}
    \centering
    \hbox{
        \includegraphics[width=\columnwidth]{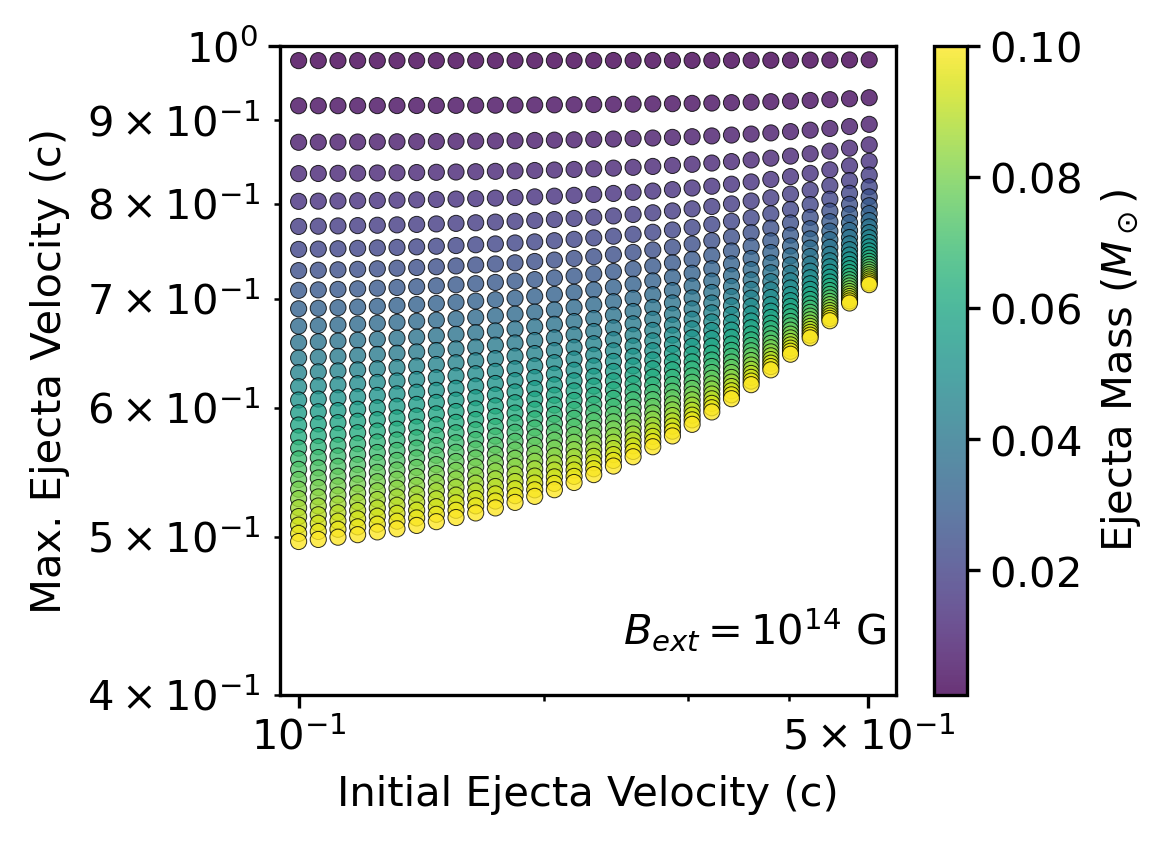}
        \includegraphics[width=\columnwidth]{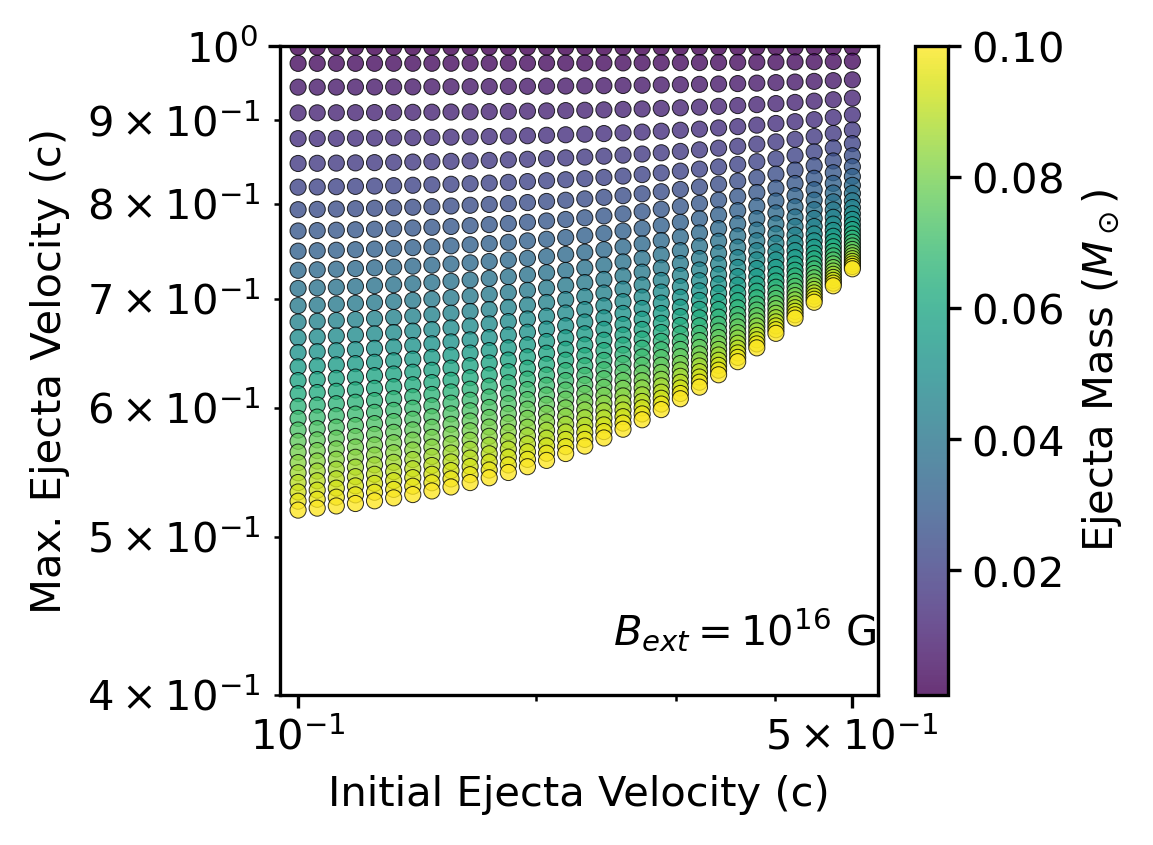} 
    }
    \caption{Maximum (dynamical or disk wind) ejecta velocities reached for a representative BNS in our simulation with energy injection from the least magnetized ($B_{\rm ext} = B_{\rm int} = 10^{14}$ G; left panel) and most magnetized ($B_{\text{\rm int}}=\frac{1}{\sqrt{3}}\times 10^{17}$\,G, $B_{\text{\rm ext}} = 1\times10^{16}$\,G; right panel) magnetar remnants considered here. For this plot we set $p_0 = 7\times10^{-4}$ s, $R_r=11$\,km, $I=3\times10^{45}$ g\,cm$^2$, $\epsilon_e=0.1$, $\epsilon_B=0.01$, $p=2.5$, $n_{\rm ISM} =  10^{-2}\text{ cm}^{-3}$ \citep{Sarin_2021}.
    \label{fig:max_beta_no_magnetar}}
\end{figure*}

\subsection{Kilonova Afterglows with Energy Injection}
\label{sec:Methods_MagnetarModels}
BNSs producing NSs as merger remnants (see Figure \ref{fig:collapse_percentages}) could be associated with kilonova radio afterglows that may be brighter than in the case of prompt collapses to BHs if the NSs are highly magnetized (magnetars). Indeed, a magnetar remnant can pump energy into the kilonova ejecta. To model this scenario, we follow the formulation by \citet{Sarin_2022} which is implemented in the \texttt{Redback} code. This formulation considers a magnetar that loses energy due to both GW and EM emission. The GW emission is related to the prolate deformation induced on the magnetar by the internal magnetic field, with ellipticity given by 
\citep{Sarin_2022}:
\begin{equation} \label{Eq:ellipticity}
\epsilon_b = 3\times 10^{-4} \left(\frac{B_{\rm int}}{B_{\rm ext}}\right)^2 B_{\rm ext,16}^2,
\end{equation}
where $B_{\rm ext}$ and $B_{\rm int}$ are the external and internal magnetic fields respectively, and the subscript $16$ indicates that the field strength expressed in units of $10^{16}$\,G. In our study, the external dipole magnetic field is assumed to have a uniform magnitude distribution extending from $10^{14}$\,G to $10^{16}$\,G, so as to span a realistic range \citep{Olausen_2014,Lu_2014, Gao_2017, Suwa_2015}. We then set a maximum ellipticity of 0.01, which implies a maximum internal magnetic field of $\frac{1}{\sqrt{3}}\times10^{17}$\,G.

The GW luminosity is calculated as \citep{Sarin_2022}:
\begin{equation}
    L_{\rm GW} \propto I^2\epsilon_b^2\Omega^6
\end{equation}
where $\Omega$ is the time-dependent spin frequency, and the moment of inertia $I$ is calculated assuming a prolate uniformly dense star:
\begin{equation}\label{Eq:MOI}
I=\frac{2}{5}M_r R_r^2(1-\epsilon_b),
\end{equation}
where $M_r$ and $R_r$ are the mass and radius of the remnant, respectively. 

The  EM luminosity is calculated assuming vacuum magnetic dipole radiation \citep{Spitkovsky_2006}:
\begin{equation}
    \label{eq:luminosity}
    L_{\rm EM} \propto B_{\rm ext}^2R^6 \Omega ^4,
\end{equation}
where $\Omega$ is the time-dependent spin frequency, $R$ is the NS radius, and $c$ is the speed of light (see also Table~\ref{tab:redbackinputs}).

In the presence of a magnetar remnant, the BNS ejecta accelerates through the magnetar's magnetic field to a maximum velocity, which is determined by both the ejecta's mass and initial velocity (Figure\,\ref{fig:max_beta_no_magnetar}). \citet{Hotokezaka_2013} place upper limits on the realistic maximum velocity of the kilonova ejecta to be $\approx 0.8c$.  Because $\approx 99.5\%$ of the BNS mergers in our 1-yr population reach maximum dynamical ejecta speeds $\gtrsim 0.8c$  even with the weakest magnetar injection (central panel in Figure\,\ref{fig:max_beta_no_magnetar}), we consider a magnetar injection scenario only for the disk wind. We further limit the simulations to disk wind cases in which the maximum ejecta speed remains $\lesssim 0.8c$. The BNSs that satisfy this last condition are about 50\% of those indicated in column 7 of Table\,\ref{tab:GW_results}.

\begin{figure}
\centering\vbox{\includegraphics[width=\columnwidth]{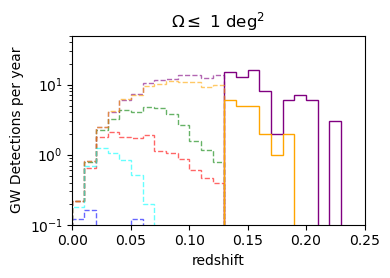} 
\vspace{-0.3cm}
\includegraphics[width=\columnwidth]{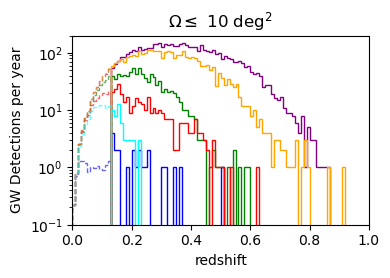} 
\vspace{-0.3cm}
\includegraphics[width=\columnwidth]{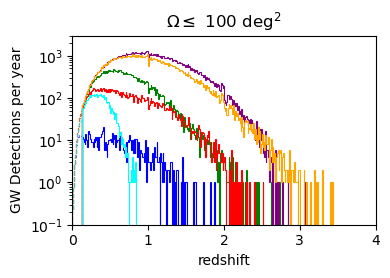} 
\vspace{-0.3cm}
\includegraphics[width=\columnwidth]{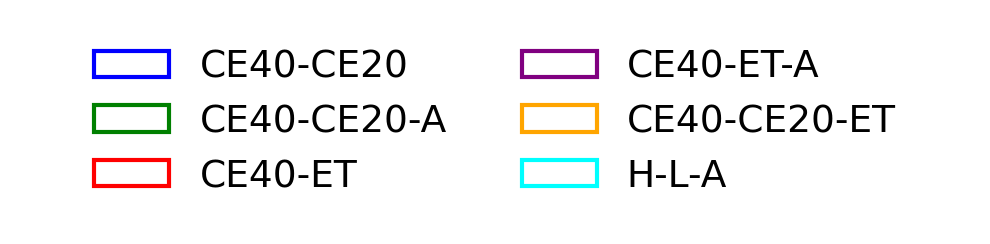}
    }
\caption{GW detections made by networks in our study, for different sky localization uncertainties and as a function redshift $z$.  
The dashed lines represent the yearly detection rate for our 50-year simulation, and the solid lines represent the number of detections made in our 1-year sample (see text for discussion). All of the \gls{BNS} mergers have aligned spins. These results are consistent with \citet{2025ApJ...985L..17P} (see their Figure 4).
\label{fig:sky_uncertainty_histograms}}
\end{figure}

\section{Results} \label{sec:results}
\subsection{GW Detection and Localization Results} \label{subsec:GWDectionResults}
We summarize our results in Table \ref{tab:GW_results} and Figure\,\ref{fig:sky_uncertainty_histograms}. 
A network of at least two widely-separated detectors is key for localization purposes. Indeed, while the yearly BNS detection rate of the H-L-A network is $\approx 400\times$ less than that of CE40-CE20---a network comprised of two US-based next-generation GW detectors, the H-L-A detection rate of well-localized events is comparable (for GW localizations $\lesssim 100$\,deg$^2$) or better (for GW localizations $\lesssim 10$\,deg$^2$) than that of CE40-CE20 (compare the cyan and blue lines in Figure\,\ref{fig:sky_uncertainty_histograms}). However, for events localized to $\lesssim 100$\,deg$^2$, the CE40-CE20 network probes BNSs at higher redshifts (median $z_{50\%}\approx 0.44$) compared to the H-L-A network ($z_{50\%}\approx 0.18$). A network of two widely separated next-generation GW detectors, such as CE40-ET (red lines in Figure\,\ref{fig:sky_uncertainty_histograms}), surpasses a three-detector network of current-generation widely-separated detectors like the H-L-A in terms of rate of GW detections with localizations $\lesssim 100$\,deg$^2$.
Networks of three widely-separated detectors with at least two of next generation (CE40-CE20-A, CE40-CE20-ET or CE40-ET-A; green, orange, and purple lines in Figure\,\ref{fig:sky_uncertainty_histograms}, respectively) are clearly superior in terms of rates of events localized to $\lesssim 10$\,deg$^2$ and ability to reach redshifts encompassing the bulk of the observed population of short GRBs, that has a median redshift of $z_{50\%}=0.68^{+0.83}_{-0.32}$ \citep{2022ApJ...940...57N}. In fact, by these metrics (yearly detection rate and maximum $z$ of well-localized events) the CE40-ET-A network, a three-detector network that maximizes the distance between detectors but contains only two next-generation instruments, outperforms the CE40-CE20-ET network, a configuration of three next-generation detectors where two (CE40 and CE20) are US-based (and hence located closer to each other).

In summary, widely separated GW detector networks are essential for building samples of well-localized BNS mergers accessible to US-based radio telescopes, with minimum (assuming a BNS local merger rate about an order of magnitude smaller than the reference value of 320\,Gpc$^{-3}$\,yr$^{-1}$)  yearly detection rates ranging from a few with current GW detectors to hundreds with next-generation GW instruments. As evident from the central panel of Figure\,\ref{fig:sky_uncertainty_histograms}, current GW networks limit well-localized detections to $z\lesssim 0.2$, while next-generation GW detectors extend the reach to $z\lesssim 0.8$. The GW network that leads to the highest number of BNSs with $SNR\ge 10$, $\Omega \le 10$\,deg$^2$, $M_{\rm tot}\le 3\,M_{\odot}$, Dec\,$>-40$\,deg (column 5 in Table \ref{tab:GW_results}) is \gls{40ETA}, in line with the highest priority recommendation given by \citet{ngGWMPSACREport}. 

Based on these results, hereafter we focus our discussion on the \gls{40ETA} network, and compare its performance with the current-generation H-L-A network.

\begin{deluxetable*}{cclllll}
\tablecaption{Summary of the properties of the BNS mergers considered in this study. Column 1: GW detector network. Column 2: yearly number of GW detections. Column 3: yearly number (fraction of BNSs in the second column) of BNS mergers with $SNR>10$ and $\Omega \le10$\,deg$^2$. Column 4: yearly number of BNS mergers with properties as in column 3 that also satisfy $M_{\rm tot}\le3\,M_{\odot}$. Column 5: yearly number of BNS mergers with properties as in column 4 that also satisfy Dec\,$\ge-40$\,deg. These constraints define the BNS sample used in our study to assess the detectability of GRB jet radio afterglows. Column 6: yearly number of BNS mergers with properties as in column 5 that also satisfy $0.7\le q \le 1$. These constraints define the BNS sample used in our study to assess the detectability of radio kilonova afterglows. Column 7: yearly number of BNS mergers with properties as in column 6 that also satisfy $M_{\rm tot}\le2.17\,M_{\odot}$. These constraints highlight well-localized BNS mergers that may leave behind stable NS remnants.  All numbers in this Table are estimated using our 1-yr BNS population, except for the ones in the last column that use our 700-yr BNS population. \label{tab:GW_results}}
\tablecolumns{7}
\tablehead{
\colhead{Network} & \colhead{$SNR\ge 10$} &  \colhead{$\Omega \leq 10$\,deg$^2$}&   \colhead{$M_{\rm tot}\le 3\,M_{\odot}$} & \colhead{Dec\,$\geq-40$\,deg} & \colhead{$0.7\le q \le 1$ }& \colhead{$M_{\rm tot}\le 2.17\,M_{\odot}$}}
\startdata
\gls{HLA} & $8.5\times10^2$   & $1.3\times10^2$ (16\%) & $7.2\times 10^1$ & $5.2\times 10^1$ & $5.1\times 10^1$& $1.4\times 10^{-3}$\\
\gls{40ETA} & $3.1\times10^5$    & $5.7\times10^3$ (1.7\%) & $3.0\times 10^3$ & $2.1\times10^3$ & $2.1\times 10^3$&  $5.9\times10^{-2}$\\
\gls{40ET} & $3.1\times10^5$  &   $4.5\times10^2$ (0.15\%)  & $2.5\times 10^2$ & $1.7 \times 10^2$ & $1.7\times 10^2$ & $5.7\times 10^{-3}$ \\
\gls{4020A} & $3.7\times10^5$  &   $1.1\times10^3$ (0.29\%) & $5.5\times10^2$ & $4.9\times 10^2$ & $4.8\times10^2$ & $1.9\times 10^{-2}$ \\
\gls{4020} & $3.6\times10^5$  &  $2.9\times10$ (0.0079\%) & $1.3\times 10^1$ & $1.0\times 10^1$ & $1.0\times 10^1$ & $0$\\
\gls{4020ET} & $4.1\times10^5$  &  $3.6\times10^3$ (0.88\%)  & $1.8\times 10^3$ & $1.4\times 10 ^3$ &$1.3\times 10^3$ & $4.0\times10^{-2}$\\
\enddata
\end{deluxetable*}
 
\subsection{GRB Jet Afterglow Results} \label{subsec:Jet_results}
Our results are summarized in Tables \ref{tab:ResultsSummary} and \ref{tab:detectable_times}.  Using the ngVLA in survey (pointed) mode at 27\,GHz (93\,GHz), short GRB-like jet afterglows can be detected for $\gtrsim 30\%$ ($\gtrsim 70\%$) of the H-L-A/CE40-20-ET well-localized BNS samples (see column 5 in Table \ref{tab:GW_results}). These lower limits are calculated by multiplying the GW\,\% indicated in column 2 of Table \ref{tab:ResultsSummary} (i.e., 49--81\% at 27\,GHz with the ngVLA in survey mode, and 95--100\% at 93\,GHz with the ngVLA in pointed mode), by the most conservative estimates for the fraction of light curves unaffected by SSA, namely, 57--60\% at 27\,GHz and 95--100\% at 93\,GHz.  

\begin{figure*}
    \centering
\hbox{
\includegraphics[width=\columnwidth]{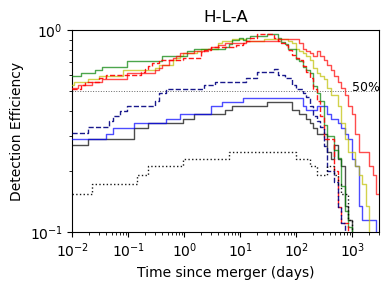}
\includegraphics[width=\columnwidth]{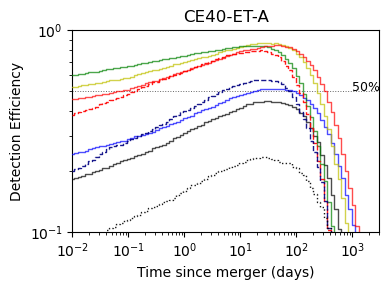}
}
\includegraphics[width=.9\textwidth]{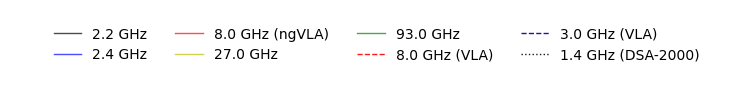}
\caption{Jet afterglow detection efficiency as a function of time for the BNSs in our 1-yr population that pass the criteria summarized in Table \ref{tab:GW_results} (column 5). Detectability is determined by the flux density falling above the $5\times$\,rms threshold for the corresponding detector and observing frequency $\nu_{\rm obs}$ (see Table~\ref{tab:ngVLAsensitivity}) and by the condition $\nu_{\rm obs} > \nu_{\rm SSA}$, where $\nu_{\rm SSA}$ is computed following the formulation  by \citet{2000ApJ...543...66P}. The detection efficiency is derived by normalizing the number of radio detections (at the corresponding time and observing frequency) by the total  number of radio detections that would be possible without considering the effects of synchrotron self-absorption (derived dividing column 2 by column 3 in Table \ref{tab:detectable_times} for the ngVLA; see also column 2 in Table \ref{tab:ResultsSummary}). Solid lines indicate ngVLA detections in pointed mode except for the 2.2\,GHz which refers to the survey mode; dashed lines indicate VLA detections; and the dotted line represents DSA-2000 detections. 
\label{fig:detectable_times}}
\end{figure*}

Considering the above results, the size of the yearly sample of well-localized BNS mergers observable by US-based radio arrays provided by the H-L-A ($\approx 52$\,yr$^{-1}$) and 40-20-ET ($\approx 2.1\times10^3$\,yr$^{-1}$) networks, and the uncertainties in the local BNS merger rate, it is clear that the CE40-ET-A plus ngVLA tandem is essential 
for providing a census of the BNSs that are able to form successful relativistic jets. The importance of this fact cannot be overstated: linking directly GRBs to their progenitors is something that,  over the past fifty years of GRB science, we have been able to do only once with GW170817. Establishing this link systematically would shed light on which BNSs are able to produce successful jets, and what are their properties. This, in turn, should help constrain the physics behind jet formation and acceleration in GRB central engines. Moreover, having radio detections extending across a wide range of radio frequencies (from 2\,GHz to 93\,GHz with the ngVLA) and timescales (see Figure \ref{fig:detectable_times}) is key for constraining the microphysics of the jet and, specifically, the power-law index of the electron energy distribution behind the shock front as related to shock-acceleration physics \citep[][]{Sari_1998,10.1093/mnras/stz2248}. We note that observations at the lowest radio frequencies (1.4---2.4\,GHz) with the ngVLA, the DSA-2000, or the VLA are feasible, though subject to synchrotron self-absorption effects (column 3 in Table \ref{tab:ResultsSummary} and Figure \ref{fig:detectable_times}). 

\begin{figure*} 
\begin{center}
\includegraphics[width=0.8\textwidth]{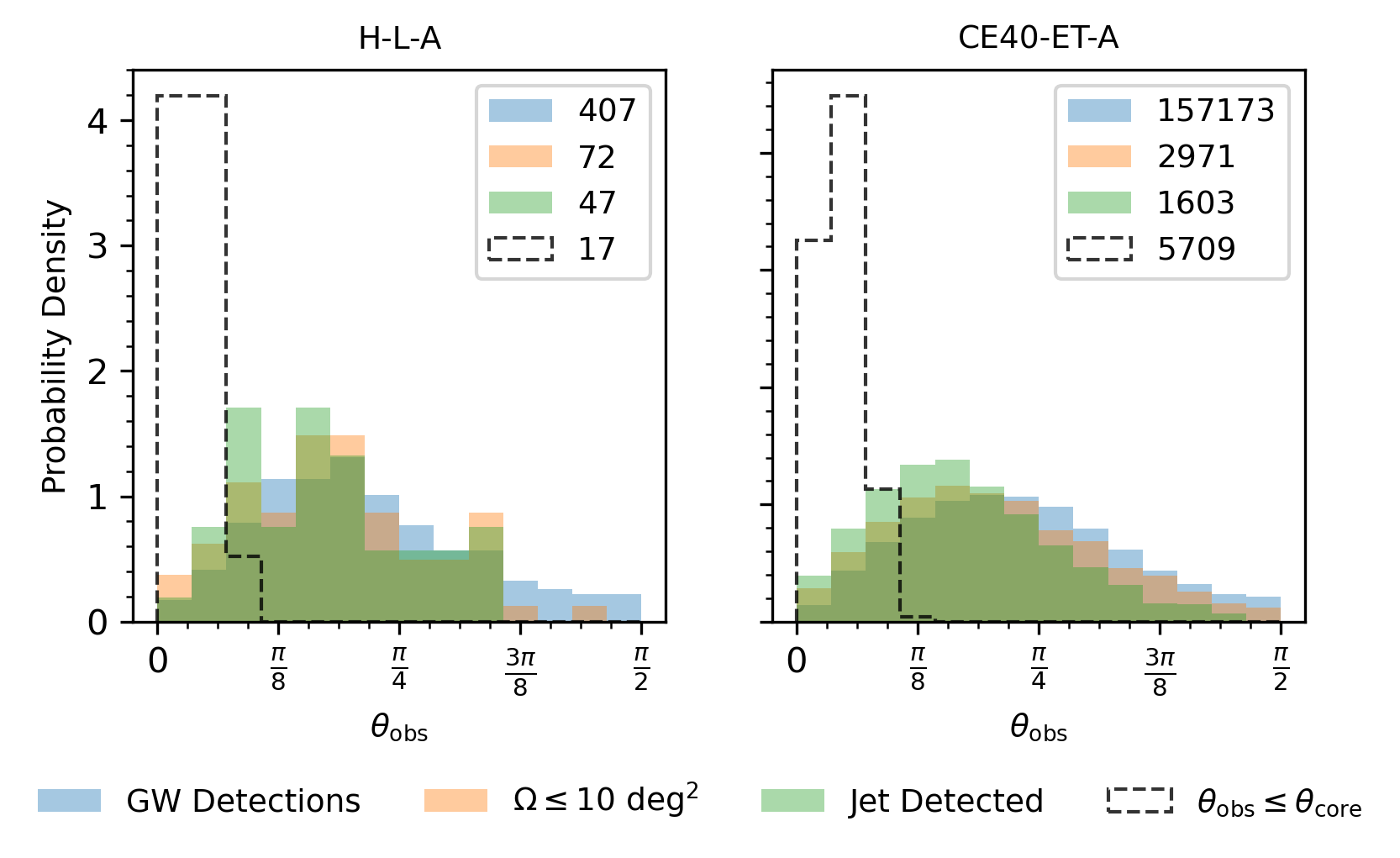}
\caption{Histograms of the observing angles of all GW detections (blue), those with the sky area less than 10 deg$^2$ (orange), and those for which jet afterglows are detected with the ngVLA at 93\,GHz (see Table~\ref{tab:ngVLAsensitivity}).\label{fig:thetaObs}}
\end{center}
\end{figure*}

In Figure~\ref{fig:thetaObs} we compare the distribution of observing angles for BNSs with $SNR\ge10$ (i.e., all GW detections, blue histograms), with that of well-localized BNSs ($SNR\ge10$ and $\Omega \le 10$\,deg$^2$, orange histograms) and of BNSs with jet afterglow detected by the ngVLA at 93\,GHz (green histograms). For detections made by \gls{40ETA} (right panel of Figure~\ref{fig:thetaObs}), the median values of the observing angles $\theta_{\rm obs}$ for all GW detections (with $M_\text{tot}\leq3M_\odot$), well-localized GW detections ($\Omega \le 10$\,deg$^2$), and detections for which a jet is also observed are $40^\circ$, $35^\circ$, and $30^\circ$ respectively. 
Hence, even though radio detections favor slightly smaller inclination angles compared to GW detections, the multi-messenger detections still probe a population dominated by off-axis events. To emphasize this point, in Figure~\ref{fig:thetaObs} we also show with black dashed lines the observing angle distribution for all BNSs in our 1-yr sample with $SNR\ge10$, $M_{\text{tot}}\le 3M_\odot$, and $\theta_{\rm obs} \le \theta_{\rm core}$ i.e., jets with their cores aligned with our line of sight and hence 
likely to be accompanied by a short-GRB-like $\gamma$-ray signals \citep[though see e.g. ][for a recent review of the main open questions regarding the prompt emission mechanisms]{2020FrASS...7...78L}. As it is evident comparing the black-dashed histograms with the green ones, radio follow ups of well-localized GW events open the possibility to identify a population of BNS jets at larger observing angles (i.e., beyond the jet cores) compared to GRBs discovered using $\gamma$-rays only. This, as demonstrated in the case of GW170817, can enable a systematic study of jet structures (namely, how the energy and speed of the shocked jet material varies as a function of polar angle). 

\begin{figure}
    \centering
\vbox{
\includegraphics[width=\columnwidth]{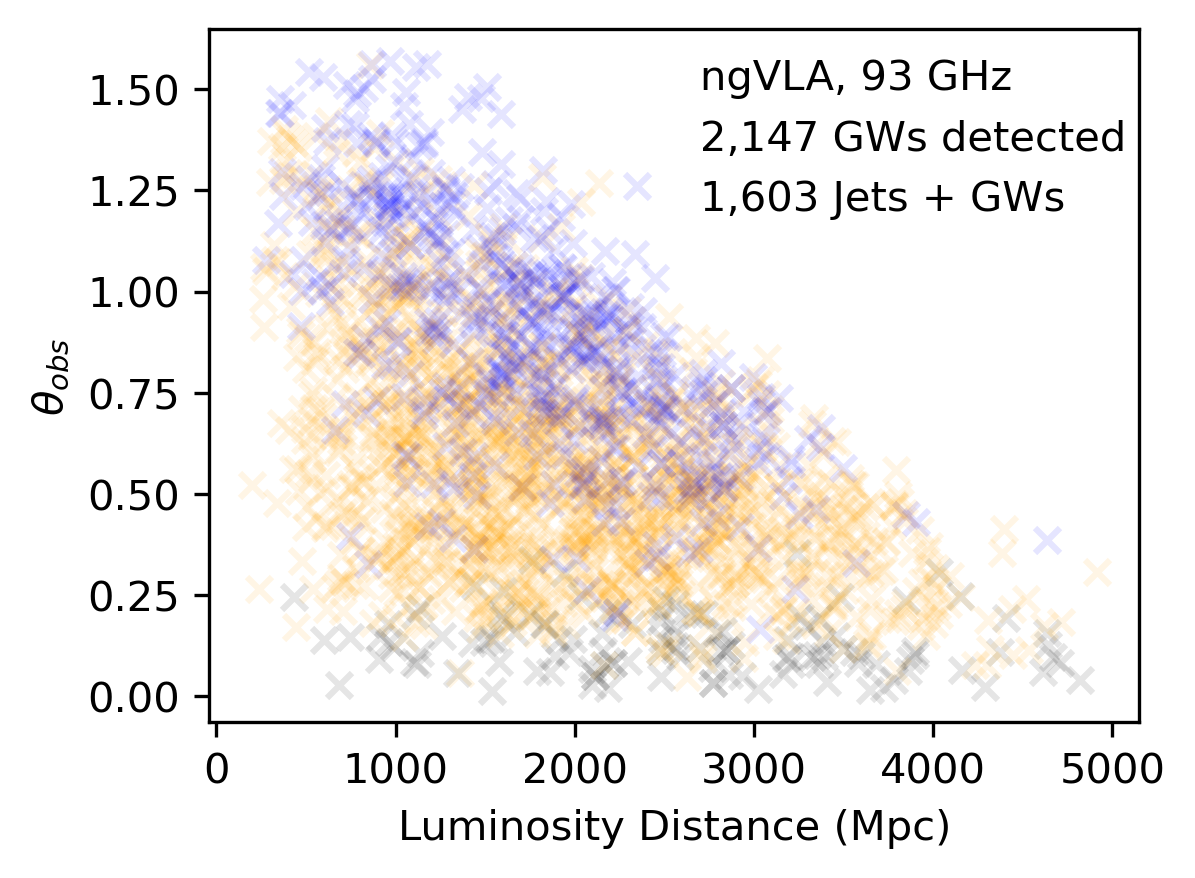}
\includegraphics[scale = 0.7]{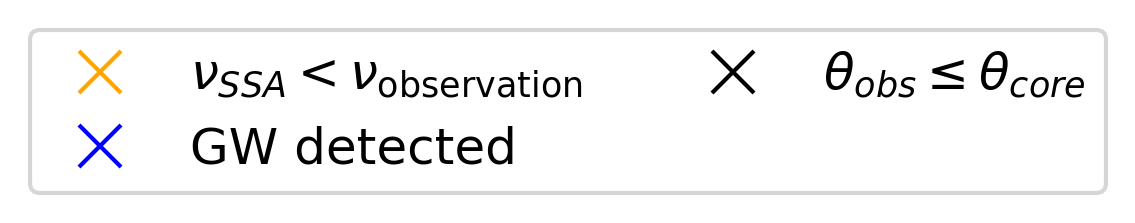}
}
\caption{Luminosity distances versus observing angles of BNSs for which GWs are detected by \gls{40ETA} (blue; 2,147 GW detections in total). In orange we plot the sub-sample of BNSs whose jet afterglows are also detected by the ngVLA at 93\,GHz with sub-milli-arcsec resolution (4.5\,$\mu$Jy rms). In black we mark the GW plus radio detections for on-axis jets.  
\label{fig:DL_thetaObs}}
\end{figure}

In Figure \ref{fig:DL_thetaObs} we show the distribution of observing angles for GRB jet afterglows that could be detected with the ngVLA at 93\,GHz in images at 0.1\,mas resolution. BNS mergers observed at $\theta_{\rm obs} > \theta_{\rm core}$ and whose radio emission centroid can be tracked over time with sub-mas resolution are candidates for radio proper motion measurements. As shown in the case of GW170817 \citep{2018Natur.561..355M}, detecting superluminal motion can directly probe the existence of a successful relativistic jet core, and break the degeneracy between jet structure and observer's angle, aiding studies aimed at constraining the Hubble constant \citep[when paired with GW standard siren measurements;][]{2018Natur.561..355M} and/or the structure of magnetic field behind relativistic shocks \citep[when paired with radio polarization measurements;][]{2018ApJ...861L..10C}.

\begin{table*}
\begin{center}
\caption{We report the number of radio detections per year for each ejecta type, using our 1-yr BNS population (Table \ref{tab:original_stat}). We also give in parenthesis the percentage of radio detections relative to the number of GW-detected and localized BNSs reported in column 5 (for GRB jets) and column 6 (for dynamical ejecta and disk winds) of Table\,\ref{tab:GW_results}. For the GRB jet afterglow case, we include a range for the fraction afterglows likely to be unaffected by SSA effects given various estimates of $\nu_{\rm SSA}$ (see Section \ref{sec:grbjet}). Because of the uncertainties on the local BNS merger rate, the number of yearly detections reported here could be up to about an order of magnitude lower (Section\,\ref{subsec:BNSMergerPopulation}). 
\label{tab:ResultsSummary}}
\begin{tabular}{llclll}
\hline\hline
Network  & \#GRBJet (GW\%) & \%GRBJet &  \#Dyn.Ej. (GW\%) & \#DiskW. (GW\%)   & \#Magn. (GW\% ) \\
 & (yr$^{-1}$)  &  $\nu_{\rm obs} \ge \nu_{\rm SSA}$ &   (yr$^{-1}$) &  (yr$^{-1}$)   &  (yr$^{-1}$) \\
\hline
\hline
 \multicolumn{6}{c}{\bf VLA [3\,GHz pointed]}\\
\gls{HLA} & $4.5\times 10^1$ (87\%)& $0\%-64\%$ & 0 (0\%)& $1.9\times 10^1$ (37\%)& $2.8\times 10^1$ (55\%)\\
\gls{40ETA} & $1.5\times10^3$ (70\%)& $0.66\%-68\%$ & 2 (0.095\%) & $1.5\times 10^2$ (6.9\%) & $1.1\times10^3$ (50\%)\\
\hline
\multicolumn{6}{c}{\bf VLA [8\,GHz pointed]}\\
\gls{HLA} & $4.5\times 10^1$ (87\%) & $11\%-98\%$ & 0 (0\%) & $1.7\times 10^1$ (33\%) & $2.8\times 10^1$ (55\%)\\
\gls{40ETA} & $1.5\times 10^3$ (68\%) & $6.6\%-95\%$ & 2 (0.094\%) & $1.0\times 10^2$ (4.8\%) & $1.0\times 10^3$ (49\%)\\
\hline
\multicolumn{6}{c}{\bf ngVLA [2.4\,GHz pointed]}\\
\gls{HLA} & $5.2\times 10^1$ (100\%) & $1.9\%-52\%$ & $2.7\times 10^1$ (53\%)& $4.3\times 10^1$ (84\%)& $2.8\times 10^1$ (55\%)\\
\gls{40ETA} & $2.0\times 10^3$ (94\%) & $1.0\%-58\%$ & $2.3\times 10^2$ (11\%) & $1.1\times 10^3$ (52\%) & $1.1\times 10^3$ (54\%)\\
\hline
\multicolumn{6}{c}{\bf ngVLA [8\,GHz pointed]}\\
\gls{HLA} & $5.2\times 10^1$ (100\%) & $12\%-96\%$ & $2.4\times 10^1$ (47\%)& $4.1\times 10^1$ (80\%) & $2.8\times 10^1$ (55\%)\\
\gls{40ETA} & $2.0\times 10^3$ (93\%) & $8.3\%-94\%$ & $1.6\times 10^2$ (7.6\%) & $9.7\times 10^2$ (46\%)& $1.1\times 10^3$ (53\%)\\
\hline
\multicolumn{6}{c}{\bf ngVLA [93\,GHz pointed]}\\
\gls{HLA} & $4.7\times 10^1$ (90\%) & $100\%$ & 2 (3.9\%)& $1.9\times 10^1$ (37\%) & $2.8\times 10^1$ (55\%)\\
\gls{40ETA} & $1.6\times 10^3$ (75\%) & $95\%-100\%$ & 6 (0.28\%) & $1.8\times 10^2$ (8.3\%) & $1.1\times 10^3$ (50\%)\\
\hline
\multicolumn{6}{c}{\bf DSA-2000 [1.4\,GHz survey]}\\
\gls{HLA} & $5.2\times 10^1$ (100\%) & $0\%-29\%$ & $1.9\times 10^1$ (37\%)& $3.6\times 10^1$ (71\%)& $2.8\times 10^1$ (55\%)\\
\gls{40ETA} & $1.9\times 10^3$ (89\%) & $0.16\%-28\%$ & $6.7\times 10^1$ (3.2\%)& $7.3\times 10^2$ (34\%) & $1.1\times 10^3$ (53\%)\\
\hline
\multicolumn{6}{c}{\bf ngVLA [2.2\,GHz survey]}\\
\gls{HLA} & $5.2\times 10^1$ (100\%) & $1.9\%-48\%$ & $1.2\times 10^1$ (24\%)& $3.3\times 10^1$ (65\%) & $2.8\times 10^1$ (55\%)\\
\gls{40ETA} & $1.9\times 10^3$ (88\%) & $0.64\%-53\%$ & $4.4\times 10^1$ (2.1\%) & $6.0\times 10^2$ (28\%) & $1.1\times 10^3$ (53\%)\\
\hline
\multicolumn{6}{c}{\bf ngVLA [27\,GHz survey]}\\
\gls{HLA} & $4.2\times 10^1$ (81\%) & $57\%-100\%$ & 0 (0\%) & 7 (14\%)& $2.6\times 10^1$ (51\%)\\
\gls{40ETA} & $1.0\times 10^3$ (49\%) & $60\%-100\%$ & 0 (0\%) & $1.1\times 10^1$ (0.52\%) & $7.8\times 10^2$ (37\%)\\
\hline
\end{tabular}
\end{center}
\end{table*}

\begin{figure*}
    \begin{center}
\hbox{\hspace{0.1\textwidth}\includegraphics[width=0.8\columnwidth]{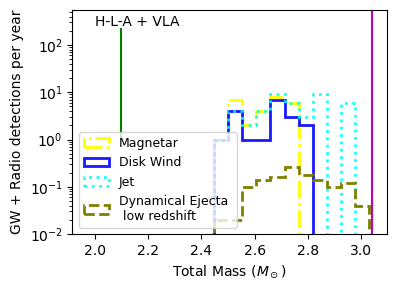}
\includegraphics[width=0.8\columnwidth]{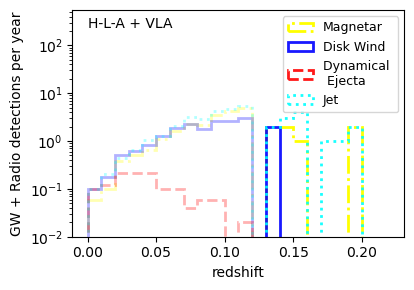}
}
\hbox{\hspace{0.1\textwidth}
\includegraphics[width=0.8\columnwidth]{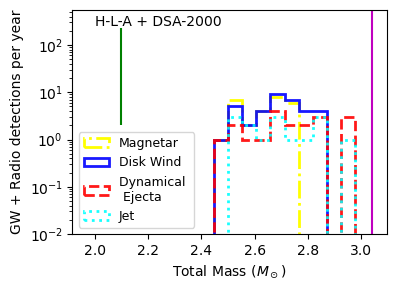}
\includegraphics[width=0.8\columnwidth]{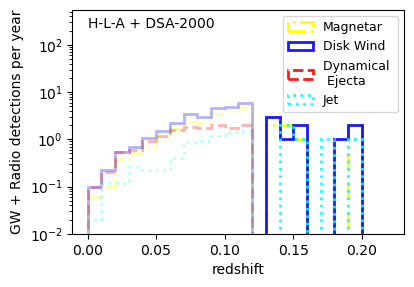}
}
\hbox{\hspace{0.1\textwidth}
\includegraphics[width=0.8\columnwidth]{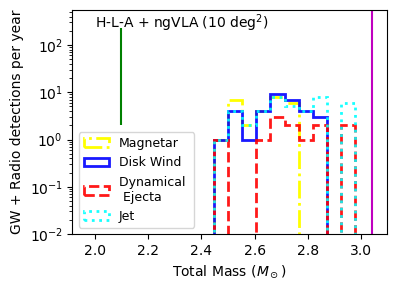}
\includegraphics[width=0.8\columnwidth]{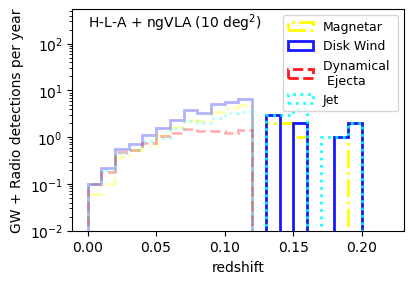}
}
\hbox{\hspace{0.1\textwidth}
\includegraphics[width=0.8\columnwidth]{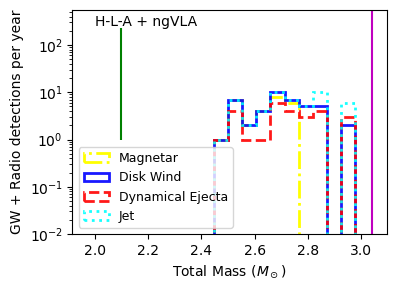}
\includegraphics[width=0.8\columnwidth]{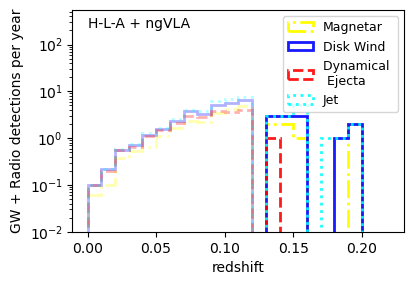}
}
\end{center}
\vspace{-1cm}
\caption{Predictions for the GW plus radio average detection rates of BNS mergers by the \gls{HLA} network and various radio arrays (as indicated at the top of each panel). When counting detections, we only consider observations made at $\nu_{\rm obs} > \nu_{\rm SSA}$ \citep{2000ApJ...543...66P}. For the ngVLA in survey mode we consider observations at 27 GHz or 2.2\,GHz. An average detection rate of 0.01\,yr$^{-1}$ implies a $\approx 10\%$ discovery (i.e. at least one detection) probability over a nominal 10-yr GW network run. Because of the uncertainties on the local BNS merger rate, the average rates shown here could be up to about an order of magnitude lower.  The solid lines are average rates derived from our 1-year population; semi-transparent lines are averages derived from the 50-yr population; green-dashed lines are averages derived from the 50-yr sample at $z\le 0.13$ (only for the VLA case, because there are no detections of the dynamical ejecta in our 700-yr sample). Left: The total mass distribution. We mark the $M_{\rm TOV}$ (green line) and $M_{\rm thr}$ (magenta line) for a $1.6\,M_{\odot}$ NS with a 12\,km  radius. Right: The redshift distribution. }
\label{fig:Mtot_Z_HLA.png}
\end{figure*}

\begin{figure*}
    \centering
\hbox{\hspace{0.1\textwidth}
\includegraphics[width=0.8\columnwidth]{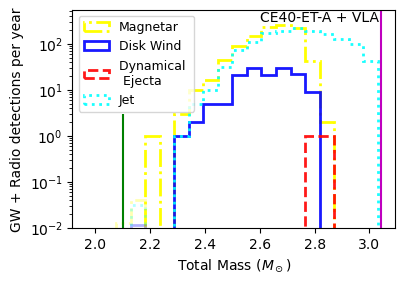}
\includegraphics[width=0.8\columnwidth]{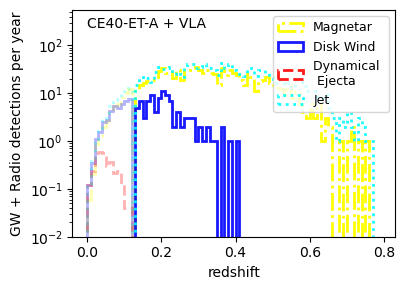}
}
\hbox{\hspace{0.1\textwidth}
\includegraphics[width=0.8\columnwidth]{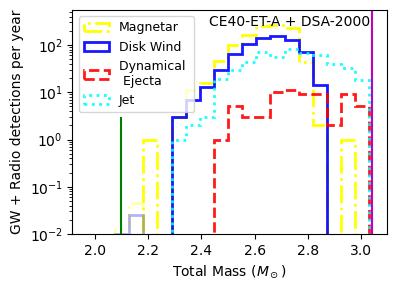}
\includegraphics[width=0.8\columnwidth]{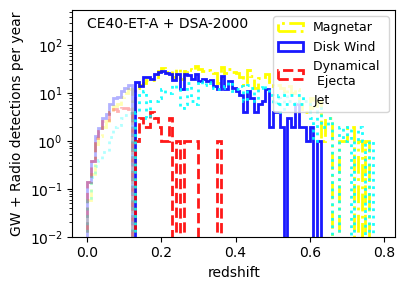}
}
\hbox{\hspace{0.1\textwidth}
\includegraphics[width=0.8\columnwidth]{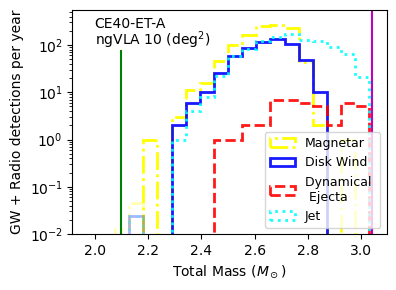}
\includegraphics[width=0.8\columnwidth]{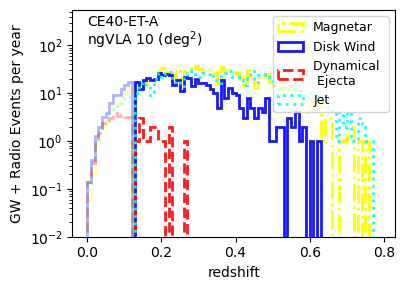}
}
\hbox{\hspace{0.1\textwidth}
\includegraphics[width=0.8\columnwidth]{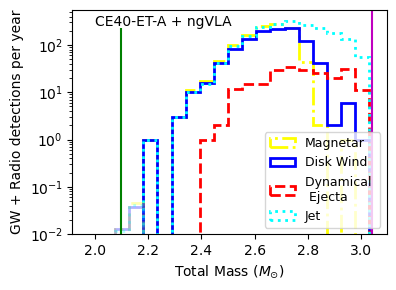}
\includegraphics[width=0.8\columnwidth]{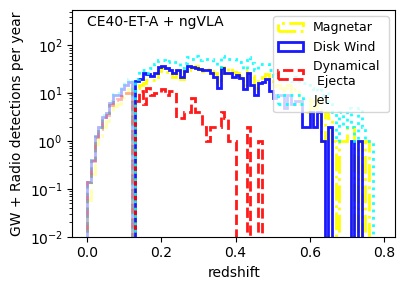}
}
\caption{Same as in Figure \ref{fig:Mtot_Z_HLA.png} but for the CE40-ET-A network. }
\label{fig:Mtot_Z_40ETA}
\end{figure*}

Finally, in Table\,\ref{tab:ResultsSummary} we compare the ngVLA performance with that of the VLA and DSA-2000, and in Figures\,\ref{fig:Mtot_Z_HLA.png}--\ref{fig:Mtot_Z_40ETA} we specifically address the ability of each GW detector network and radio array combination to probe the diversity of BNS mergers in terms of their total mass and redshift distributions. As shown by the cyan distributions in Figures \ref{fig:Mtot_Z_HLA.png}--\ref{fig:Mtot_Z_40ETA}, the distribution of jet afterglow detections (at any frequency) across $M_{\rm tot}$ and $z$ is dominated by the GW detectability rather than by the sensitivity of the radio arrays. This is evident when considering that all radio arrays have pretty high detection efficiency for GRB jet afterglows at frequencies unaffected by SSA  (see the column 2 in Table \ref{tab:ResultsSummary}). As discussed earlier in this Section, the yearly number of GW plus radio (at any frequency) detections is dominated by both the GW detectability and SSA effects. 

Overall, the CE-40-ET and ngVLA combination stands out as the most promising pairing for multi-messenger detection of GRB afterglows. This tandem not only offers high annual detection rates, but also enables uniquely rich scientific returns from these events, surpassing what is achievable with any other detector configurations considered in this study.

\begin{figure*}
    \centering
    \hbox{
        \includegraphics[width=\columnwidth]{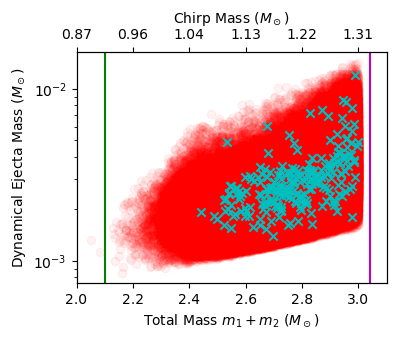}
        \includegraphics[width=\columnwidth]{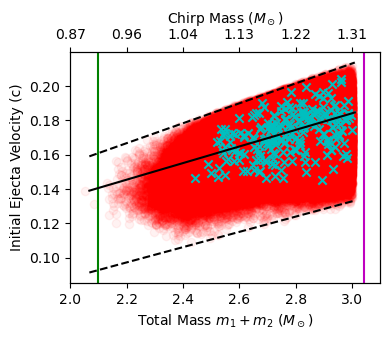}
    } 
\caption{Left: The dynamical ejecta mass of each of the BNS mergers in our 1-year simulation (red). Right: The initial ejecta velocity of the dynamical ejecta (red). We overlay the dynamical ejecta velocity with the maximum and minimum fits given by \citet{Coughlin_2019} (black solid and dashed lines). In both panels, the bottom horizontal axis shows the total mass, and the top horizontal axis shows the corresponding chirp mass for a system with $q=1$. We mark the $M_{\rm TOV}$ (green line) and $M_{\rm thr}$ (magenta line) for a $1.6\,M_{\odot}$ NS with a 12\,km  radius \citep{Margalit_2019}. 
\label{fig:velocity_initial_of_DynEj_detections}}
\end{figure*}

\begin{figure*}
    \centering
    \hbox{
        \includegraphics[width=\columnwidth]{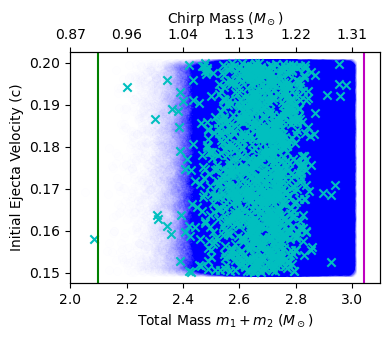} \includegraphics[width=\columnwidth]{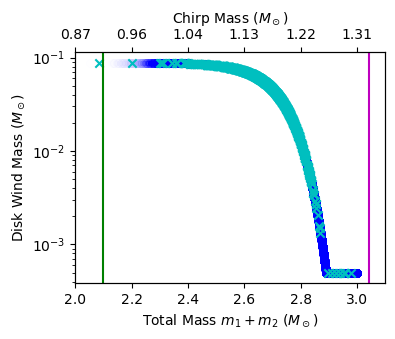} 
    }
\caption{ Same as in Figure \ref{fig:velocity_initial_of_DynEj_detections} but for the disk wind. See text for discussion. 
\label{fig:total_ejecta_mass_wDetections_diskwindB0}}
\end{figure*}

\begin{figure*}
    \centering
\vbox{
\hbox{
\includegraphics[width=\columnwidth]{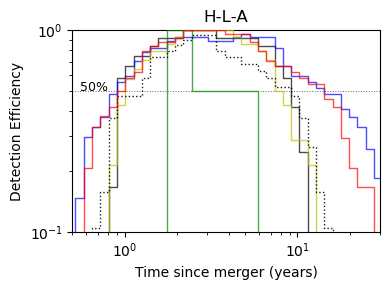}
\includegraphics[width=\columnwidth]{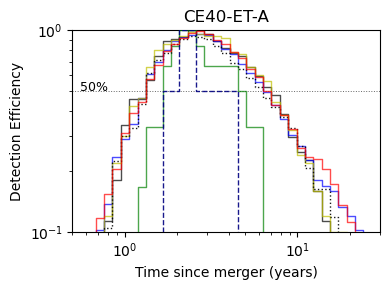}
}
\includegraphics[width=.9\textwidth]{Jet_det_efficiency_legend.png}}
\caption{Dynamical ejecta afterglow detection efficiency as a function of time for the BNSs in our 1-yr population that pass the criteria summarized in Table \ref{tab:GW_results} (column 6). Detectability is determined by the flux density falling above the $5\times$\,rms threshold for the corresponding detector and observing frequency $\nu_{\rm obs}$ (see Table~\ref{tab:ngVLAsensitivity}). The detection efficiency is derived by normalizing the number of radio detections (at the corresponding time and observing frequency) by the total  number of radio detections at that frequency (see column 3 of Table~\ref{tab:detectable_times_DynEj} for the ngVLA). Solid lines indicate ngVLA detections with the ngVLA in pointed mode except for the 2.2\,GHz which refers to the survey mode; dashed lines indicate VLA detections; and the dotted line represents DSA-2000 detections. 
\label{fig:Detectable_Times_DynEj}}
\end{figure*}

\begin{figure*}
    \centering
\vbox{
\hbox{
\includegraphics[width=\columnwidth]{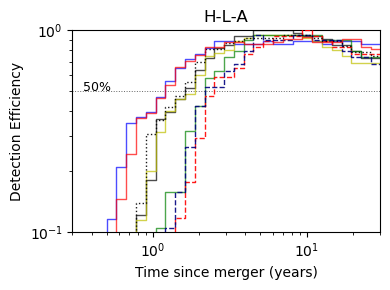}
\includegraphics[width=\columnwidth]{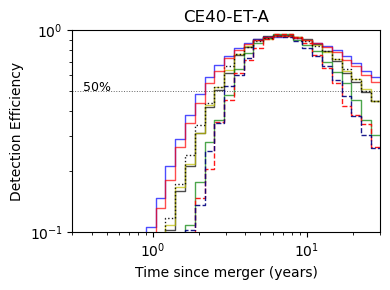}
}
\includegraphics[width=.9\textwidth]{Jet_det_efficiency_legend.png}}
\caption{Disk wind afterglow detection efficiency as a function of time for the BNSs in our 1-yr population that pass the criteria summarized in Table \ref{tab:GW_results} (column 6). Detectability is determined by the flux density falling above the $5\times$\,rms threshold for the corresponding detector and observing frequency $\nu_{\rm obs}$ (see Table~\ref{tab:ngVLAsensitivity}). The detection efficiency is derived by normalizing the number of radio detections (at the corresponding time and observing frequency) by the total  number of radio detections at that frequency (see column 3 of Table~\ref{tab:detectable_times_DiskWind} for the ngVLA). Solid lines indicate ngVLA detections with the ngVLA in pointed mode except for the 2.2\,GHz which refers to the survey mode; dashed lines indicate VLA detections; and the dotted line represents DSA-2000 detections. 
\label{fig:Detectable_Times_DiskW}}
\end{figure*}

\subsection{Kilonova Afterglow Results}
Our results are summarized in Tables \ref{tab:ResultsSummary} and \ref{tab:detectable_times_DynEj}--\ref{tab:detectable_times_DiskWind}. As evident from these Tables, in the case of the kilonova ejecta radio detection efficiencies are much lower than in the case of jet afterglows. Indeed, as evident from columns 4--5 in Table \ref{tab:ResultsSummary}, the ngVLA detection efficiency for all GW detector networks is lower than the jet afterglow case, and lowest for the dynamical ejecta afterglow ($\lesssim 50\%$ in all cases). Hence, for the kilonova ejecta, the radio detectors' sensitivities impact substantially the multi-messenger yearly detection rates, as well as the $M_{\rm tot}$ and $z$ distribution of the BNS mergers that can be studied via multi-messenger observations. This can be seen in 
Figures \ref{fig:Mtot_Z_HLA.png}-\ref{fig:Mtot_Z_40ETA}. As evident from these Figures, while the VLA can start probing the disk wind afterglow in the nearby universe and for the higher mass BNSs, the ngVLA or DSA-2000 sensitivities are needed to probe the dynamical ejecta afterglows, as the last remain out of reach at VLA sensitivities with both the current generation (H-L-A) and next generation (CE40-ET-A) GW detector networks. The ngVLA or DSA-2000, paired with an H-L-A network, can probe a sizable number ($\mathcal{O}(10)$ per year; Table \ref{tab:ResultsSummary}) of dynamical ejecta afterglows and an even larger number of disk wind afterglows, but only within the relatively nearby universe ($z\lesssim 0.2$). The CE40-ET-A network working with the ngVLA can extend multi-messenger detections of dynamical ejecta afterglows to $z\lesssim 0.4$.

In Figures~\ref{fig:velocity_initial_of_DynEj_detections}, we plot the dynamical ejecta masses (left) and initial ejecta velocities (right) of BNSs in our simulation with $SNR\ge 10$, $\Omega \le 10$\,deg$^2$, $M_{\rm tot}\le 3\,M_{\odot}$, Dec\,$>-40$\,deg, $0.7\le q \le 1$ (column 6 in Table \ref{tab:GW_results}). We overplot in cyan corresponding values for the BNSs in the sample for which radio detections are possible with the ngVLA. We present similar plots, but for the disk wind component, in Figure \ref{fig:total_ejecta_mass_wDetections_diskwindB0}. As evident from these Figures, in the case of the dynamical ejecta, larger values of $M_{\rm tot}$ and higher values of ejecta velocities favor multi-messenger detections.  In the case of the disk wind, the high-mass systems are easier to detect in GWs, but the mass of the disk wind drops off sharply at higher masses making the a radio detection more difficult. For the disk wind, the range of examined initial ejecta velocities is small enough that radio detection rates are nearly independent of it.

Finally, in Figures~\ref{fig:Detectable_Times_DynEj} and \ref{fig:Detectable_Times_DiskW}, we show the kilonova radio afterglow detection efficiency at each radio frequency and as a function of time since merger, normalized by the total number of radio detections at each frequency. The highest detection efficiencies occur at the lower radio frequencies (see Tables \ref{tab:detectable_times_DynEj} and \ref{tab:detectable_times_DiskWind}). At all frequencies, both the dynamical ejecta and disk wind components of the kilonova radio afterglows remain detectable for several years. 

\begin{figure*}
    \hbox{
        \includegraphics[width=\columnwidth]{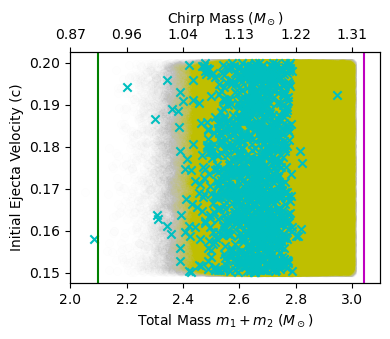} \\
        \includegraphics[width=\columnwidth]{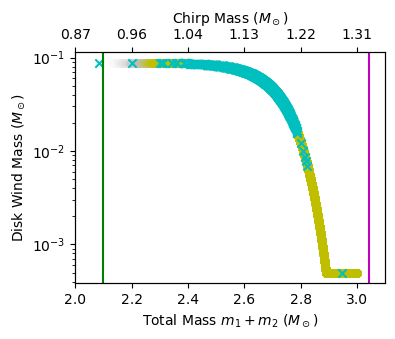} \\
    }
\caption{Same as in Figure \ref{fig:velocity_initial_of_DynEj_detections} but for the disk wind with magentar injection. See text for discussion. 
\label{fig:total_ejecta_mass_wDetections_magnetar}}
\end{figure*}

\begin{figure*}
    \centering
\hbox{
\includegraphics[width=\columnwidth]{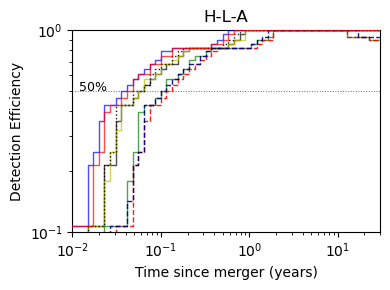}
\includegraphics[width=\columnwidth]{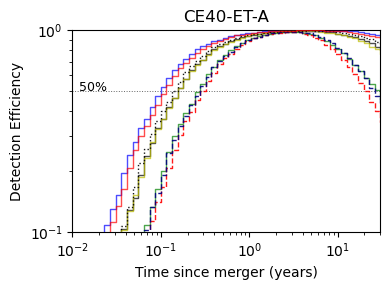}
}

\includegraphics[width=.9\textwidth]{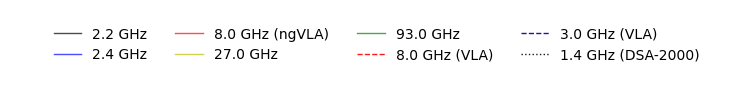}
\caption{Same as Figure\,\ref{fig:detectable_times} but for the disk wind with magnetar injection.  At each frequency and for each radio array, the detection efficiency is derived by normalizing the number of radio detections at the corresponding time by the total  number of radio detections at that frequency (see column 3 of Table~\ref{tab:detectable_times_redback} for the ngVLA). 
\label{fig:detectable_times_redback}}
\end{figure*}

\subsubsection{Kilonova Afterglow with Energy Injection Results} \label{subsubsec:RadioResultsMagnetar}

Our results are summarized in Tables \ref{tab:ResultsSummary} and  \ref{tab:detectable_times_redback}. In the case of a magnetar central engine, as expected, the wind disk afterglow is much brighter. Roughly 50\% of BNSs with $SNR\ge 10$, $\Omega \le 10$\,deg$^2$, $M_{\rm tot}\le 3\,M_{\odot}$, Dec\,$>-40$\,deg, $0.7\le q \le 1$ (column 7 in Table\,\ref{tab:GW_results}) maintain a maximum ejecta velocity $\le 0.8c$. Hence, the $\approx 50\%$ radio detection efficiencies reported in column 6 of Table \ref{tab:ResultsSummary} almost uniformly across the considered radio arrays, simply reflect the fact that any radio array can detect the bright magnetar-powered disk winds, if a magnetar is indeed the merger remnant. This implies that even with the current-generation GW detectors at their improved sensitivities (H-L-A network) and the VLA, we should start probing these systems systematically. However, because of the limited sensitivity achievable by current GW detectors, studies with the H-L-A network can only probe magnetar-powered kilonova afterglows at $z\lesssim 0.2$, while the CE40-ET-A network can extend these detections to $z\lesssim 0.8$ and to the lower $M_{\rm tot}$ range (hence probing BNSs systems that could form stable NSs).  

In Figure~\ref{fig:total_ejecta_mass_wDetections_magnetar}, we plot the disk wind masses (left) and initial velocities (right) of BNSs in our simulation with $SNR\ge 10$, $\Omega \le 10$\,deg$^2$, $M_{\rm tot}\le 3\,M_{\odot}$, Dec\,$>-40$\,deg, $0.7\le q \le 1$, and maximum ejecta velocities $\lesssim 0.8c$. We overplot in cyan corresponding values for the BNSs in the sample for which radio detections are possible with the ngVLA. We see no significant relation between the initial ejecta velocity and the radio detectability. But we see that there are only a few detections with $M_{\rm tot} \ge 2.8\,M_{\odot}$. This is because BNSs with larger total masses would create less massive disk winds that can accelerate to velocities $\>0.8c$ when a magnetar injects energy into the ejecta. 

Finally, in Figure~\ref{fig:detectable_times_redback}, we show the radio afterglow detection efficiency for the disk wind with magnetar injection at each radio frequency and as a function of time since merger, normalized by the total number of radio detections at each frequency. As in the jet afterglow case, the highest detection efficiencies occur at the lower radio frequencies (see Tables \ref{tab:detectable_times_DynEj} and \ref{tab:detectable_times_DiskWind}). At all frequencies, this component remains detectable for several years. 

\section{Summary and Conclusion}\label{sec:conclusion}
We have presented a study aimed at quantifying the multi-messenger detection prospects of BNS mergers with current and next generation GW detectors and radio arrays, focusing specifically on well-localized GW events and BNSs with total masses $\lesssim 3$\,M$_{\odot}$. To account for current uncertainties in radio afterglow models, we have considered a variety of potential radio counterparts including the afterglows of relativistic jets, kilonova dynamical and disk wind ejecta, and disk wind ejecta with a magnetar central engine. Our results can be summarized as follows:
\begin{itemize}
\item {\it GW detections and localizations:} Having a network of at least two widely-separated detectors is key for localization purposes. The H-L-A detection rate of well-localized events (GW localizations $\lesssim 10$\,deg$^2$) surpasses that of CE40-CE20. Networks of three widely-separated detectors with at least two of next generation (CE40-CE20-A, CE40-CE20-ET or CE40-ET-A) are clearly superior in terms of rates of events localized to $\lesssim 10$\,deg$^2$ and ability to reach redshifts encompassing the observed short GRB population. These results are in full agreement with the recommendations by \citet{ngGWMPSACREport} and results by \citet{2025ApJ...985L..17P}. 

\item {\it Jet afterglows:} For BNSs that are well-localized ($\Omega \le 10$\,deg$^2$) in GWs, the prospects for detecting associated radio jets, if those jets are formed, are very promising with radio arrays operating at frequencies that are unaffected by SSA. The $M_{\rm tot}$ and $z$ distribution of BNSs for which we can achieve multi-messenger detections via their jet afterglows is dominated by the GW detectability rather than by the sensitivity of the radio arrays. 
The improved sensitivity of the CE40-ET-A network (compared to the H-L-A) boosts yearly detection rates of BNS mergers and extend these detections to redshifts encompassing the median redshift of the observed short GRB population. Overall, the CE-40-ET and ngVLA combination stands out as the most promising pairing for multi-messenger detection of GRB afterglows. This tandem not only offers high annual detection rates, but also enables uniquely rich scientific returns from these events, surpassing what is achievable with any other detector configurations considered in this study.
\item {\it Kilonova ejecta afterglows:} While the VLA can start probing the disk wind afterglow in the nearby universe and for the higher mass BNSs, ngVLA or DSA-2000 sensitivities are needed to probe the dynamical ejecta afterglows. The last, differently from the disk wind afterglows,  remain out of reach at VLA sensitivities with both the current generation (H-L-A) and next generation (CE40-ET-A) GW detector networks. With the ngVLA or DSA-2000 and an H-L-A network, it will be possible to probe kilonova ejecta afterglows within the relatively nearby universe ($z\lesssim 0.2$). The CE40-ET-A network working with the ngVLA and/or DSA-2000 can extend multi-messenger detections of these afterglows to $z\lesssim 0.4-0.8$ and to lower values of $M_{\rm tot}$ where quasi-stable NSs may form as merger remnants. Exploring the dependence of the kilonova afterglow properties (relative importance of disk versus wind component, and ejecta speed distribution) on the nature of the merger remnant is likely to shed light on the astrophysics of the ejecta while helping to constrain the NS EoS \citep[e.g.,][and references therein]{Nedora_2021b,Balasubramanian_2022}.

\item {\it Disk wind afterglows with magnetar remnants:}  Magnetar-driven disk wind radio afterglows are so bright that a detection is guaranteed even with the VLA for all BNSs with good GW localizations. Hence, even with the current-generation GW detectors at their improved sensitivities (H-L-A network), we should start probing these systems systematically. Next generation detector networks like CE40-ET-A can extend these detections to $z\lesssim 0.8$ and to the lower $M_{\rm tot}$ range (hence probing BNSs systems that could form stable NSs), even with the VLA.  
\end{itemize}

In conclusion, as evident from the above summary, the next decade will offer a unique opportunity to fundamentally transform the field of multi-messenger astronomy. It will be possible to transition from an era of rare detections, to one where a flood of detections enable BNS population studies, while opening new phase space for exciting discoveries. 

\begin{acknowledgments}
K.M. and A.C. acknowledge support from the National Science Foundation and from the US Department of Energy. This work was partially carried out at the Advanced Research Computing at Hopkins (ARCH) core facility  (rockfish.jhu.edu). We thank colleagues of the Cosmic Explorer Management Team and of the LIGO Scientific Collaboration for providing comments that improved this manuscript. This research has made use of data obtained from the Gravitational Wave Open Science Center (gwosc.org), a service of the LIGO Scientific Collaboration, the Virgo Collaboration, and KAGRA.
\end{acknowledgments}

\appendix
\begin{figure}
\includegraphics[width=\columnwidth]{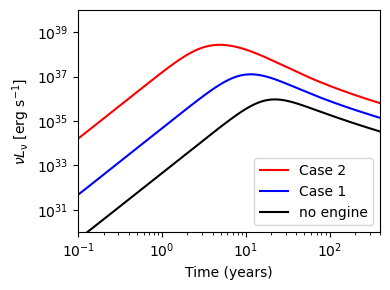}
\caption{We replicate the kilonova afterglow radio light curves shown in  Figure 8 of \cite{Sarin_2022} for two specific magnetar cases. Case 1 represents a magnetar with $B_{\rm int} = 10^{16.5}$\,G and $B_{\rm ext}=10^{14.5}$\,G. Case 2 describes a magnetar with $B_{\rm int} = B_{\rm ext} = 10^{16}$\,G. 
\label{fig:Figure8}}
\end{figure}
\section{Generating Light Curves with Redback} \label{sec:Appendix}
As a way of checking the consistency of our methods with previous work \citep{Sarin_2022}, we replicated the radio portion (1\,GHz light curves) of Figure 8 in \citet{Sarin_2022}. Our version is shown in Figure~\ref{fig:Figure8} and utilizes the following parameters: $B_{\rm int}=10^{16.5}$\,G and $B_{\rm ext}=10^{14.5}$\, for Case 1; $B_{\rm int}=B_{\rm int}=10^{16}$\,G for Case 2; $p_0=7\times10^{-4}$\,s; $R_r=11$\,km; $I=3\times10^{45}$\,g\,cm$^{2}$; $n_{\rm ISM}=0.01$\,g\,cm$^{-3}$; $z=0.1$; ejecta mass of 0.05\,$M_{\odot}$. The other parameters are set as indicated in Table\,\ref{tab:redbackinputs}.

The process for running the \texttt{Redback} code to generate light curves is three-fold. First, we compute the time-dependent magnetar luminosity which pumps energy into the kilonova ejecta. We do this using the \texttt{magnetar\_luminosity\_evolution} function of \texttt{Redback}'s \texttt{transient\_models.magnetar\_models} module which returns an array of luminosities calculated at each input time.  This array of luminosities becomes an input for the \texttt{\_ejecta\_dynamics\_and\_interaction()} function.\footnote{See \texttt{Redback}'s \texttt{transient\_models.magnetar\_driven\_ejecta\_models} module.} This function returns an output kinetic energy tuple (given by \texttt{output.kinetic\_energy}) whose maximum is taken as initial kinetic energy into \texttt{kilonova\_afterglow\_redback()}. The last then outputs a flux density in units of mJy.

\section{Radio Detection Tables}
\begin{deluxetable}{cll}
\tablecaption{Number of GRB jet afterglows per year detectable by the various GW networks and the ngVLA. At each frequency, we assume the sensitivities reported in Table~\ref{tab:ngVLAsensitivity} for the pointed mode observations, except at 2.2\,GHz for which we assume the sensitivity in survey mode. We record only the detections for which $\nu_{\text{SSA}}<\nu_{\text{obs}}$, assuming the most optimistic value for $\nu_{\rm SSA}$ \citep[][corresponding to the highest percentages reported in column 3 of Table \ref{tab:ResultsSummary}]{2000ApJ...543...66P}. In parenthesis we give the percentage of radio detections relative to the number of BNSs reported in column 5 of Table \ref{tab:GW_results}. 
\label{tab:detectable_times}}
\tablecolumns{3}
\tablewidth{0pt}
\tablehead{
\colhead{Network} & \colhead{Freq. } & \colhead{\#GRBJet (GW\%)} \\
 & \colhead{(GHz)} & (yr$^{-1}$) }
\startdata
\gls{HLA} & 2.2 & $2.5\times10^1$ (48\%) \\ 
 & 2.4 & $2.7\times10^1$ (52\%) \\
 & 8.0 & $5.0\times10^1$ (96\%) \\
 & 16 & $5.2\times10^1$ (100\%) \\
 & 27 & $5.2\times10^1$ (100\%) \\
 & 41 & $4.9\times10^1$ (94\%) \\
 & 93 & $4.7\times10^1$ (90\%) \\
\gls{40ETA} & 2.2  & $1.0\times 10^3$ (46\%) \\
 & 2.4 & $1.2\times 10^3$ (55\%) \\
 & 8.0 & $1.9\times 10^3$ (88\%) \\
 & 16 & $1.9\times 10^3$ (90\%) \\
 & 27 & $1.9\times 10^3$ (88\%) \\
 & 41 & $1.8\times 10^3$ (85\%) \\
 & 93 & $1.6\times 10^3$ (75\%) 
\enddata
\end{deluxetable}

\begin{deluxetable}{cll}
\tablecaption{Number of dynamical ejecta afterglows per year detectable by the various GW networks and the ngVLA. At each frequency, we assume the sensitivities reported in Table~\ref{tab:ngVLAsensitivity} for the pointed mode observations, except at 2.2\,GHz for which we assume the sensitivity in survey mode. In parenthesis we give the percentage of radio detections relative to the number of BNSs reported in column 6 of Table \ref{tab:GW_results}.  \label{tab:detectable_times_DynEj}}
\tablecolumns{5}
\tablewidth{0pt}

\tablehead{
\colhead{Network} & 
\colhead{Freq.} & 
\colhead{\#Dyn.Ej. (GW\%)}\\
& 
\colhead{(GHz)} & (yr$^{-1}$) 
}
\startdata
\gls{HLA} & 2.2 & $1.2\times 10^1$ (24\%)  \\
 & 2.4 & $2.7\times 10^1$ (53\%)  \\
 & 8.0 & $2.4\times 10^1$ (47\%)  \\
 & 16 & $2.0\times 10^1$ (39\%)   \\
 & 27 & $1.4\times 10^1$ (27\%)  \\
 & 41 & $1.1\times 10^1$ (22\%)   \\
 & 93 & $2.0\times 10^0$ (3.9\%) \\
 \gls{40ETA} & 2.2 & $4.4\times 10^1$ (2.1\%) \\
 & 2.4 & $2.3\times 10^2$ (11\%)  \\
 & 8.0 & $1.6\times 10^2$ (7.6\%)  \\
 & 16 & $8.4\times 10^1$ (4.0\%) \\
 & 27 & $5.0\times 10^1$ (2.4\%) \\
 & 41 & $3.0\times 10^1$ (1.4\%)  \\
 & 93 & $6.0\times 10^0$ (0.28\%) 
\enddata 
\end{deluxetable}

\begin{deluxetable}{cll}
\tablecaption{Number of disk wind afterglows per year detectable by the various GW networks and the ngVLA. At each frequency, we assume the sensitivities reported in Table~\ref{tab:ngVLAsensitivity} for the pointed mode observations, except at 2.2\,GHz for which we assume the sensitivity in survey mode. In parenthesis we give the percentage of radio detections relative to the number of BNSs reported in column 6 of Table \ref{tab:GW_results}.  \label{tab:detectable_times_DiskWind}}
\tablecolumns{5}
\tablewidth{0pt}

\tablehead{
\colhead{Network} & 
\colhead{Freq.} & 
\colhead{\#DiskW. (GW\%)}\\
& 
\colhead{(GHz)} & 
 (yr$^{-1}$) 
}
\startdata
\gls{HLA} & 2.2 & $3.3\times 10^1$ (65\%)  \\
 & 2.4 & $4.3\times 10^1$ (84\%)  \\
 & 8.0 & $4.1\times 10^1$ (80\%)  \\
 & 16 & $3.6\times 10^1$ (71\%)  \\
 & 27 & $3.5\times 10^1$ (69\%)   \\
 & 41 & $3.2\times 10^1$ (63\%)  \\
 & 93 & $1.9\times 10^1$ (37\%)  \\
 \gls{40ETA} & 2.2 & $6.0\times10^2$ (28\%) \\
 & 2.4 & $1.1\times 10^3$ (52\%)  \\
 & 8.0 & $9.7\times10^2$ (46\%)   \\
 & 16 & $7.5\times10^2$ (35\%)  \\
 & 27 & $6.0\times10^2$ (28\%)  \\
 & 41 & $4.5\times10^2$ (21\%)  \\
 & 93 & $1.8\times10^2$ (8.3\%) 
\enddata 
\end{deluxetable}

\begin{deluxetable}{cll}
\tablecaption{Number of disk wind afterglows per year detectable by the various GW networks and the ngVLA, assuming a magnetar central engine. In parenthesis we give the percentage of detections relative to the numbers reported in column six of Table \ref{tab:GW_results}. At each frequency, we assume the sensitivities reported in Table~\ref{tab:ngVLAsensitivity}. \label{tab:detectable_times_redback}}
\tablecolumns{5}
\tablewidth{0pt} 

\tablehead{
\colhead{Network} & 
\colhead{Freq.} & 
\colhead{\#Magn. (GW\%)}\\
& 
\colhead{(GHz)} & (yr$^{-1}$)
}
\startdata
\gls{HLA} & 2.2 & $2.8\times 10^1$ (55\%)  \\
 & 2.4 & $2.8\times 10^1$ (55\%)  \\
 & 8.0 & $2.8\times 10^1$ (55\%)  \\
 & 16 & $2.8\times 10^1$ (55\%)  \\
 & 27 & $2.8\times 10^1$ (55\%)  \\
 & 41 & $2.8\times 10^1$ (55\%)  \\
 & 93 & $2.8\times 10^1$ (55\%)  \\
\gls{40ETA} & 2.2 & $1.1\times 10^3$ (53\%)  \\
 & 2.4 & $1.1\times 10^3$ (54\%)  \\
 & 8.0 & $1.1\times 10^3$ (53\%)  \\
 & 16 & $1.1\times 10^3$ (53\%)  \\
 & 27 & $1.1\times 10^3$ (53\%)  \\
 & 41 & $1.1\times 10^3$ (52\%)  \\
 & 93 & $1.1\times 10^3$ (50\%) 
\enddata 
\end{deluxetable}

\bibliography{main}{}
\bibliographystyle{aasjournal}

\end{document}